\documentclass[%
reprint,
numerical,
prd,
superscriptaddress,
twocolumn,
 nofootinbib,
 amsmath,
 amssymb,
]{revtex4-2}

\usepackage[utf8]{inputenc}
\usepackage{graphicx}% Include figure files
\usepackage{dcolumn}% Align table columns on decimal point
\usepackage{color}
\usepackage{booktabs}
\usepackage{acronym}
\usepackage{ifthen}
\usepackage{blindtext}
\usepackage[normalem]{ulem}
\usepackage[colorlinks=true, linkcolor=blue, urlcolor=blue, citecolor=blue]{hyperref}
\usepackage{etoolbox}
\usepackage{fancyhdr}
\usepackage{xspace}
\usepackage{textcomp}
\usepackage{multirow}
\usepackage[caption=false]{subfig}
\usepackage{lineno}
\usepackage{tabularx}
\usepackage{aas_macros}
\usepackage{orcidlink}

\def\pbhs{primordial black holes\xspace}

\newacro{S/N}{signal-to-noise ratio}

\newacro{PN}{post-Newtonian}

\newacro{O3}{the third observing run}

\newacro{O2}{the second observing run}

\newacro{MSP}{millisecond pulsar}

\newacro{CW}{\emph{Continuous Wave}}

\newacro{FFT}{fast Fourier transform}
\newacro{CNN}{convolutional neural network}

\newcommand{\D}{{\mathcal{D}}}
\newcommand{\Dul}{{\mathcal{D}^{95\%}}}
\newcommand{\F}{{\mathcal F}}

\newcommand{\hul}{{h_0^{95\%}}}

\newcommand{\pf}{\textrm{P}_{\textrm{f}}}
\newcommand{\Tsft}{T_{\textrm{SFT}}}
\newcommand{\Tcoh}{T_{\textrm{coh}}}
\newcommand{\Tobs}{T_{\textrm{obs}}}

\newcommand{\soap}{{\em SOAP}}
\newcommand{\fh}{{\em FrequencyHough}}
\newcommand{\sh}{{\em SkyHough}}

\newcommand{\fstat}{{\em $\mathcal{F}$-statistic}}
\def\tdfstat{{{\em Time-Domain} \fstat}}

\newcommand{\be}{\begin{equation}}
\newcommand{\ee}{\end{equation}}
\newcommand{\bea}{\begin{eqnarray}}
\newcommand{\eea}{\end{eqnarray}}

\newtoggle{fullauthorlist}
\toggletrue{fullauthorlist}
\newtoggle{endauthorlist}
\toggletrue{endauthorlist}

\newcommand\scinum[2]{\ensuremath{#1\!\times\!10^{#2}}}

\begin{document}

%\linenumbers
\title{All-sky search for continuous gravitational waves from isolated neutron stars using Advanced LIGO and Advanced Virgo O3 data}
%\preprint{APS/123-QED}


%\iftoggle{fullauthorlist}{
%  \input{LSC-Virgo-KAGRA-Authors-Aug-2020-aas.tex}
%}

%  \author{The LIGO Scientific Collaboration}
%  \affiliation{LSC}
%  \author{The Virgo Collaboration}
%  \affiliation{Virgo}
%  \author{The KAGRA Collaboration}
%  \affiliation{KAGRA}
  
\iftoggle{endauthorlist}{
  %
  % Put the author list at the end of the document.
  % Save author, affiliation, and maketitle commands.
  %
  \let\mymaketitle\maketitle
  \let\myauthor\author
  \let\myaffiliation\affiliation
  \author{The LIGO Scientific Collaboration}
  \author{The Virgo Collaboration}
  \author{The KAGRA Collaboration}
  \email{Full author list given at the end of the article.}
\noaffiliation
}
{

%\iftoggle{fullauthorlist}{
%  \input{KAGRA-LSC-Virgo-Authors-Aug-2020-aas}
%}
%{
%  \author{The LIGO Scientific Collaboration}
%  \affiliation{LSC}
%  \author{The Virgo Collaboration}
%  \affiliation{Virgo}
%  \author{The KAGRA Collaboration}
%  \affiliation{KAGRA}  
%}
}

\date{\today}% It is always \today, today,

\begin{abstract}
We present results of an all-sky search for continuous gravitational waves which can be produced by spinning neutron stars with an asymmetry around their rotation axis, using data from the third observing run of the Advanced LIGO and Advanced Virgo detectors. Four different analysis methods are used to search in a gravitational-wave frequency band from 10 to 2048 Hz and a first frequency derivative from $-10^{-8}$ to $10^{-9}$ Hz/s. No statistically-significant periodic gravitational-wave signal is observed by any of the four searches. As a result, upper limits on the gravitational-wave strain amplitude $h_0$ are calculated. The best upper limits are obtained in the frequency range of 100 to 200 Hz and they are ${\sim}1.1\times10^{-25}$ at 95\% confidence-level. The minimum upper limit of  $1.10\times10^{-25}$ is achieved at a frequency 111.5 Hz. We also place constraints on the rates and abundances of nearby planetary- and asteroid-mass primordial black holes that could give rise to continuous gravitational-wave signals.
\end{abstract}

%\keywords{gravitational waves --- pulsars: general}

\maketitle

\section{Introduction}
\label{sect:intro}

The Advanced LIGO \cite{2015CQGra..32g4001L} and Advanced Virgo \cite{2015CQGra..32b4001A} detectors have made numerous detections of gravitational waves (GW), to date consisting of short-duration (transient) GW emitted during the inspirals and mergers of compact binary systems of black holes (BH), neutron stars (NS), \cite{PhysRevX.9.031040,2021PhRvX..11b1053A}, as well as mixed NS-BH binaries \cite{2021ApJ...915L...5A}. Among still undiscovered types of GW radiation are long-lasting, almost-monochromatic continuous waves (CW), whose amplitudes and frequencies change much more slowly compared to those of transient sources (on the timescale of years rather than seconds). Astrophysically, promising sources of CW are  rotating, non-axisymmetric NS, emitting GW at a frequency close to, or related to, their spin frequency. Deviations from the symmetry (a NS `deformation') may be caused by fluid instabilities, such as in the case of r-modes, or by elastic, thermal or magnetic stresses in the crust and/or core of NS, and may be acquired at various stages of stars' isolated evolution, or during an interaction with a companion in a binary system (for recent reviews on sources of CW, see e.g., \cite{2015PASA...32...34L,2018ASSL..457..673G,2019Univ....5..217S}). Discovery of CW emitted by NS would allow to probe their still mysterious interiors, study properties of dense matter in conditions distinct from those occurring in inspirals and mergers of binary NS systems, as well as carry out additional tests of the theory of gravity \cite{PhysRevD.96.042001}. Due to intrinsically smaller GW amplitude of CW in comparison to the already-detected transient sources, searches for CW from rotating non-axisymmetric NS are essentially limited to the Galaxy. 

The search presented here is not limited to gravitational-wave signals from deformed rotating neutron stars. Another source of quasi-monochromatic, persistent GWs are very light, planetary- and asteroid-mass, inspiraling \pbhs (PBHs), which could comprise a fraction or the totality of dark matter \cite{Green:2020jor}. Such signals would arise from inspiraling PBHs whose chirp masses are less than $\mathcal{O}(10^{-5})M_\odot$ and whose GW frequencies are less than $\sim 250$ Hz, and would be indistinguishable from those arising from non-axisymmetric rotating NSs spinning up.

Recent detections of black holes made by the LIGO-Virgo-KAGRA Collaboration have revived interest in PBHs: low spin measurements and the rate inferences are consistent with those expected for BHs that formed in the early universe \cite{Clesse:2020ghq}. Existence of light PBHs is well-motivated theoretically and experimentally: recent detections of star and quasar microlensing events~\cite{Niikura:2019kqi,Hawkins:2020zie,bhatiani2019confirmation} suggest compact objects or PBHs with masses between $10^{-6}$ and $10^{-5}~M_\odot$ could constitute a fraction of dark matter of order $f_{\rm PBH} \sim 0.01$, which is consistent within the unified scenario for PBH formation presented in ~\cite{clesse2018seven}, but greater than expected for free-floating (i.e. not bound to an orbit) planets \cite{sivaram2019primordial} (e.g. the hypothetical Planet 9 could be a PBH with a mass of $ 10^{-6} M_\odot$ that was captured by the solar system \cite{Scholtz:2019csj}). PBHs may also collide with NS and be responsible for the origin of NS-mass BHs, potentially detectable in the LIGO-Virgo-KAGRA searches \cite{2018ApJ...868...17A}. 
However, constraints arising from such observations \cite{Green:2020jor}, even those that come from the LIGO-Virgo merging rate inferences \cite{LIGOScientific:2019kan,Phukon:2021cus} and stochastic background searches \cite{Raidal:2017mfl,Wang:2016ana}, rely on modelling assumptions, and can be evaded if, for example, PBHs formed in clusters \cite{Garcia-Bellido:2017xvr,Calcino:2018mwh,Belotsky:2018wph,Carr:2019kxo,Trashorras:2020mwn,DeLuca:2020jug}. It is therefore important to develop complementary probes of these mass regimes to test different PBH formation models \cite{Miller:2020kmv,Miller:2021knj}, which is possible by searching for continuous GWs. 

Searches for continuous waves are usually split in three different domains: {\em targeted searches} look for signals from known pulsars; {\em directed searches} look for signals from known sky locations; {\em all-sky searches} look for signals from unknown sources. All-sky searches for {\em a priori} unknown CW sources have been carried out in the Advanced LIGO and Advanced Virgo data previously \cite{2017PhRvD..96f2002A,2018PhRvD..97j2003A,2017PhRvD..96l2004A,2019PhRvD.100b4004A,PhysRevLett.123.171101,2021ApJ...909...79S,2020PhRvL.125q1101D,2021PhRvD.103f3019D,PhysRevD.104.082004,Covas:2020nwy,LIGOScientific:2020qhb,2021PhRvD.103h3020W,LIGOScientific:2021jlr}. A recent review on pipelines for wide parameter-space searches can be found in~\cite{Tenorio:2021wmz}. 

Here we report on results from an all-sky, broad frequency range search using the most-sensitive data to date, the LIGO-Virgo O3 observing run, employing four different search pipelines: the {\fh} \cite{2014PhRvD..90d2002A}, {\sh} \cite{2004PhRvD..70h2001K}, {\tdfstat} \cite{1998PhRvD..58f3001J,Aasi:2014}, and {\soap} \cite{2019PhRvD.100b3006B}.
Each pipeline uses different data analysis methods and covers different regions of the frequency and frequency time derivative parameter space, although there exist overlaps between them (see Table \ref{tab:ffdot_range_summary} and Fig.~\ref{fig:ffdot_range_summary} for details). The search is performed for frequencies between 10 Hz and 2048 Hz and for a range of frequency time derivative between -$10^{-8}$ Hz/s and $10^{-9}$ Hz/s, covering the whole sky. 
We note here that the search is generally-agnostic to the type of the GW source, so the results are not actually limited to signals from non-axisymmetric rotating NS in our Galaxy. A comprehensive multi-stage analysis of the signal outliers obtained by the four pipelines has not revealed any viable candidate for a continuous GW signal. However we improve the broad-range frequency upper limits with respect to previous O1 and O2 observing run and also with respect to the recent analysis of the first half of the O3 run \cite{PhysRevD.104.082004}. This is also the first all-sky search for CW sources that uses the Advanced Virgo detector's data. 

The article is organized as follows: in Section \ref{sect:data} we describe the O3 observing run and provide details about the data used. Section \ref{sect:overview} we present an overview of the pipelines used in the search. Section \ref{sect:methods}, details of the data-analysis pipelines are described. Section \ref{sect:results}, we describe the results obtained by each pipeline, namely the signal candidates and  the sensitivity of the search whereas Section \ref{sect:conclusions} contains a discussion of the astrophysical implications of our results. 

\section{Data sets used}\label{sect:data}
The data set used in this analysis was the third observing run (O3) of the Advanced LIGO and Advanced Virgo GW detectors \citep{2015CQGra..32g4001L,2015CQGra..32b4001A}. LIGO is made up of two laser interferometers, both with 4\,km long arms. One is at the LIGO Livingston Observatory (L1) in Louisiana, USA and the other is at the LIGO Hanford Observatory (H1) in Washington, USA. Virgo (V1) consists of one interferometer with 3\,km arms located at European Gravitational Observatory (EGO) in Cascina, Italy. The O3 run took place between the 2019 April 1 and the 2020 March 27. The run was divided into two parts, O3a and O3b, separated by one month commissioning break that took place in October 2019. The duty factors for this run were $\sim76\%$, $\sim71\%$, $\sim76\%$ for L1, H1, V1 respectively. The maximum uncertainties (68\% confidence interval) on the calibration of the LIGO data were of 7\%/11\% in magnitude and 4 deg/9 deg in phase for O3a/O3b data  (\citep{2020CQGra..37v5008S, 2021arXiv210700129S}). For Virgo, it amounted to 5$\%$ in amplitude and 2 deg in phase, with the exception of the band 46 - 51 Hz, for which the maximum uncertainty was estimated as 40$\%$ in amplitude and 34 deg in phase during O3b. For the smaller range 49.5 - 50.5 Hz, the calibration was unreliable during the whole run \citep{2021arXiv210703294V}.

\section{Overview of Search Pipelines}
\label{sect:overview}

In this section we provide a broad overview of the four pipelines used in the search. The three pipelines: {\fh}, {\sh}, and {\tdfstat} have been used before in several all-sky searches of the LIGO data. The {\soap} pipeline is a new pipeline applied for the first time to an all-sky search. It uses novel algorithms. {\soap} aims at a fast, preliminary search of the data before more sensitive but much more time consuming methods are applied (see~\cite{Tenorio:2021wmz} for a review on pipelines for wide parameter-space searches).
The individual pipelines are described in more detail in the following section.

\subsection{Signal model}\label{sect:sigmodel}
The GW signal in the detector frame from an isolated, asymmetric NS spinning around one of its principal axis of inertia is given by \cite{1998PhRvD..58f3001J}:
\begin{eqnarray}
\begin{aligned}
    h(t) &=h_0[F_+(t,\alpha,\delta,\psi)\frac{1+\cos^2{\iota}}{2} \cos{\phi(t)} \\
    &+ F_{\times}(t,\alpha,\delta,\psi) \cos{\iota} \sin{\phi(t)} ],
    \label{eq:h0t}
\end{aligned}
\end{eqnarray}
where $F_+$ and $F_{\times}$ are the antenna patterns of the detectors dependent on right ascension $\alpha$, declination $\delta$ of the source and polarization angle $\psi$, $h_0$ is the amplitude of the signal, $\iota$ is the angle between the total angular momentum vector
of the star and the direction from the star to the Earth, and $\phi(t)$ is the phase of the signal. The amplitude of the signal is given by:
\begin{eqnarray}
\begin{aligned}
  \label{eqn:hexpected}
h_0 &= \frac{4\pi^2G}{c^4} \frac{\epsilon I_{zz}  f^2}{d}  \approx                  1.06\times10^{-26}\left(\frac{\epsilon}{10^{-6}}\right) \\
    &\times\left(\frac{I_{zz}}{10^{38}\>\rm{kg\ m}^2}\right)\left(\frac{f}{100\>{\rm  Hz}}\right)^2
\left(\frac{1\>{\rm kpc}}{d}\right), 
\end{aligned} 
\end{eqnarray}
where $d$ is the distance from the detector to the source, $f$ is the GW frequency (assumed to be twice the rotation frequency of the NS), $\epsilon$ is the ellipticity or asymmetry of the star, given by $(I_{xx}-I_{yy})/I_{zz}$, and $I_{zz}$ is the moment of inertia of the star with respect to the principal axis aligned with the rotation axis.

We assume that the phase evolution of the GW signal can be approximated with a second order Taylor expansion around a fiducial reference time $\tau_r$:
\begin{align}
    \phi(\tau) = \phi_o + 2\pi [f (\tau-\tau_r) + \frac{\dot{f}}{2!} (\tau-\tau_r)^2],
    \label{eq:phaseevo}
\end{align}
where $\phi_o$ is an initial phase and $f$ and $\dot{f}$ are the frequency and first frequency derivative at the reference time. The relation between the time at the source $\tau$ and the time at the detector $t$ is given by:
\begin{equation}
  \label{eq:tau(t)}
  \tau(t) = t + \frac{\vec{r}(t) \cdot \vec{n}}{c} +
  \Delta_{E\odot} - \Delta_{S\odot}\, , 
\end{equation}
where $\vec{r}(t)$ is the position vector of the detector in the Solar System Barycenter (SSB) frame, and $\vec{n}$ is the unit vector pointing to the NS; $\Delta_{E\odot}$ and $\Delta_{S\odot}$ are respectively the
relativistic Einstein and Shapiro time delays.
In standard equatorial coordinates with right ascension $\alpha$ and
declination $\delta$, the components of the unit vector $\vec{n}$ are
given by $(\mathrm{cos}\,\alpha \, \mathrm{cos}\,\delta,\
\mathrm{sin}\,\alpha \, \mathrm{cos}\,\delta,\ \mathrm{sin}
\,\delta)$.

\subsection{Parameter space analyzed}

\begin{figure} 
\hspace*{-25pt}
\includegraphics[width=1.1\columnwidth]{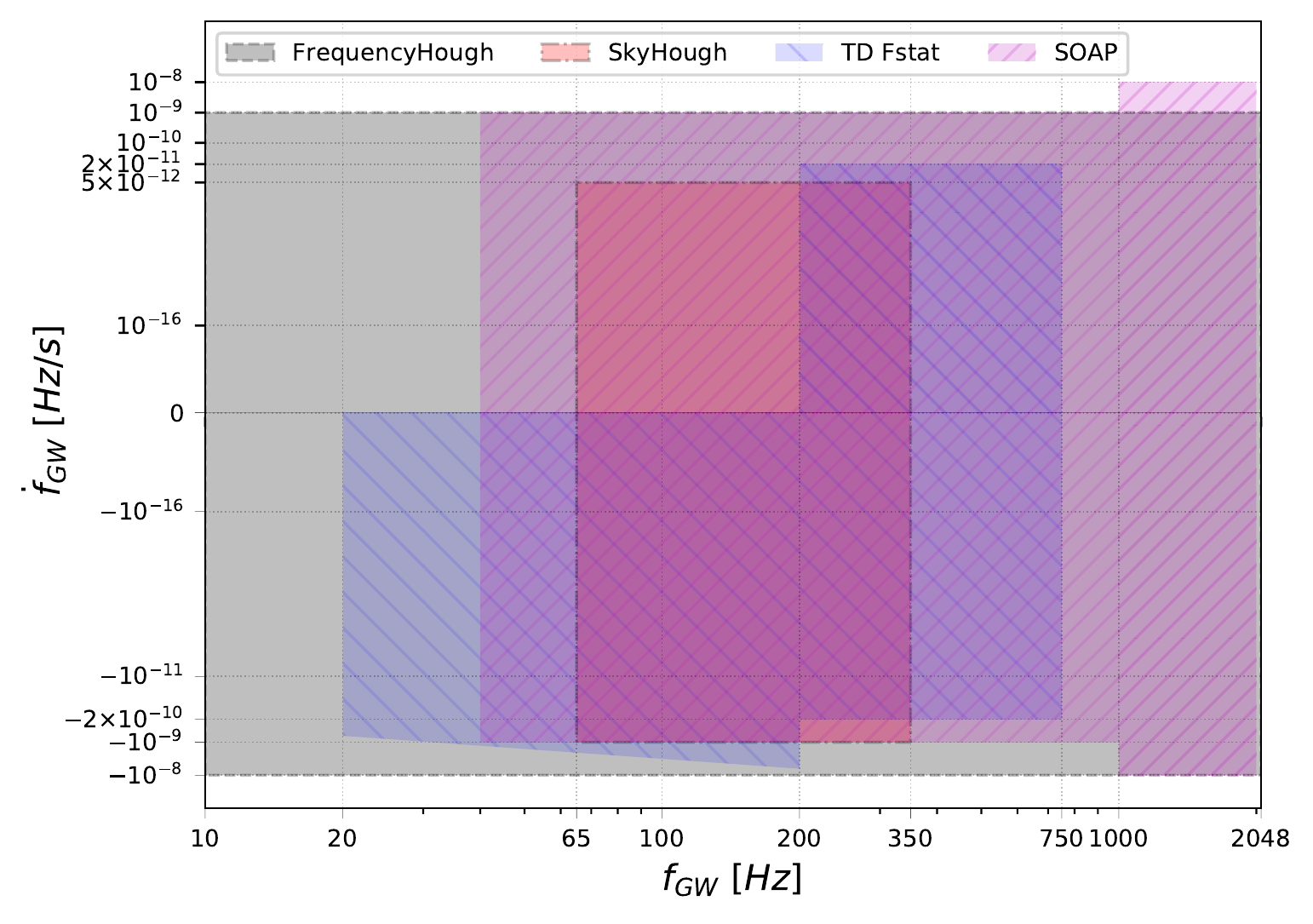}
\caption{Frequency and frequency derivative search ranges of the four pipelines: the {\fh} pipeline ranges marked in grey, {\sh} in red, {\tdfstat} in blue, and {\soap} in magenta. See Table~\ref{tab:ffdot_range_summary} for details.} 
\label{fig:ffdot_range_summary} 
\end{figure}

All the four pipelines perform an all-sky search, however the frequency and frequency derivative ranges analyzed are different for each pipeline.
The detailed ranges analyzed by the four pipelines are summarized in Table \ref{tab:ffdot_range_summary} and presented in Fig.~\ref{fig:ffdot_range_summary}.
\begin{table*}[tbp]
\begin{center}
\begin{tabular}{l l l}
\hline
Pipeline & Frequency [Hz] & Frequency derivative [Hz/s] \\
\hline \hline
FrequencyHough &  $10 - 2048$ &    -$10^{-8} - 10^{-9}$ \\
SkyHough       &  $65 - 350$ &     -$10^{-9}  - 5\times 10^{-12}$ \\                
SOAP           &  $40 - 1000$ &    -$10^{-9}  - 10^{-9}$ \\
               &  $1000 - 2000$ &  -$10^{-8}  - 10^{-8}$ \\
TD Fstat      &  $20 - 200$ &     -$3.2\times 10^{-9} f/100  - 0$ \\ 
               &  $200 - 750$ &    -$2\times 10^{-10}  - 2\times 10^{-11}$ \\
\hline
\end{tabular}
\caption[]{Frequency and frequency derivative search ranges of the four pipelines.}
\label{tab:ffdot_range_summary}
\end{center}
\end{table*}
The {\fh} pipeline analyzes a broad frequency range between 10 Hz and 2048 Hz and a broad frequency time derivative range between -$10^{-8}$ Hz/s and $10^{-9}$ Hz/s. A very similar range
of $f$ and $\dot{f}$ is analyzed by {\soap} pipeline. The {\sh} pipeline analyzes a narrower frequency range where the detectors are most sensitive whereas {\tdfstat} pipeline analyzes $f$ and $\dot{f}$ ranges of the bulk of the observed pulsar population (see Fig.~\ref{fig:ffdot_gw} in Sect. \ref{ssect:FStatm}).

\subsection{Detection statistics}

As all-sky searches cover a large parameter space they are computationally very expensive and it is computationally prohibitive to analyze coherently the data from the full observing run using optimal matched-filtering. As a result each of the pipelines developed for the analysis uses a semi-coherent method. Moreover to reduce the computer memory and to parallelize the searches the data are divided into narrow bands.
Each analysis begins with sets of {\em short Fourier transforms} (SFTs) that span the observation period, with coherence times ranging from 1024s to 8192s. The {\fh}, {\sh} and {\soap} pipelines compute measures of strain power directly from the SFTs and create detection statistics by stacking those powers with corrections for frequency evolution applied. The {\fh} and {\sh} pipelines use {\em Hough} transform
to do the stacking whereas {\soap} pipeline uses the {\em Viterbi} algorithm. 
The {\tdfstat} pipeline extracts band-limited 6-day long time-domain data segments from the SFT sets and applies frequency evolution corrections coherently to obtain the $\F$-statistic (\cite{1998PhRvD..58f3001J}). Coincidences are then required among multiple data segments with no stacking.

\subsection{Outlier follow-up}

All four pipelines perform a follow-up analysis of the statistically significant candidates (outliers) obtained during the  search. All pipelines perform vetoing of the outliers corresponding to narrow, instrumental artifacts (lines) in the advanced LIGO detectors (\cite{bib:O3Linelist}). Several other consistency vetoes are also applied to eliminate outliers. The {\fh}, {\sh}, and {\tdfstat} pipelines
perform follow-up of the candidates by processing the data with increasing long coherence times whereas {\soap} pipeline use {\em convolutional neural networks} to do the post processing. 

\subsection{Upper limits}

No periodic gravitational wave signals were observed by any of the  four pipelines and and all the pipelines obtain upper limits on their strength.  The three pipelines {\sh}, {\tdfstat} and {\soap} 
obtain the upper limits by injections of the signals according to the model given in Section \ref{sect:sigmodel} above for an array of signal amplitudes $h_0$ and randomly choosing the remaining parameters. The {\fh} pipeline obtains upper limits using an analytic formula (see Eq.\,\ref{eq:hul})
that depends on the spectral density of the noise of the detector. The formula was validated by a number of tests consisting of injecting signals to the data.

\section{Details of search methods}
\label{sect:methods}

\subsection{\fh}
\label{ssect:FHm}
The \fh\ pipeline is a semi-coherent procedure in which interesting points (i.e. outliers) are selected in the signal parameter space, and then are followed-up in order to confirm or reject them. This method has been used in several past all-sky searches of Virgo and LIGO data~\citep{2016PhRvD..93d2007A,2017PhRvD..96f2002A,2019PhRvD.100b4004A,PhysRevLett.123.171101}. A detailed description of the methodology can be found in \citep{2014PhRvD..90d2002A}.
In the following, we briefly describe the main analysis steps and specific choices used in the search.
 
Calibrated detector data are used to build ``short duration'' and cleaned~\cite{2005CQGra..22S1197A} Fast Fourier Transform (FFTs), with duration $T_{\text{FFT}}$ which depends on the frequency band being considered, see Table~\ref{tab:fh_fft}. 

Next, local maxima are selected based on the square root of the equalized power of the data 
\footnote{Computed as the ratio of the squared modulus of each FFT of the data and an auto-regressive estimation of the average power spectrum, see \cite{2005CQGra..22S1197A} for more details.}
passing a dimensionless threshold of $\Theta=1.58$. The collection of these time-frequency peaks forms the so-called {\it peakmap}.

The peakmap is cleaned of the strongest disturbances using a {\it line persistency} veto~\cite{2014PhRvD..90d2002A}.

\begin{table}[tbp]
\begin{center}
\begin{tabular}{ccccc}\hline
Band [Hz] & $T_{\text{FFT}}$ [s] &  $\delta f$ [Hz] & $\delta \dot{f}$ [Hz/s]\\
\hline \hline
  $10$--$128$ & 8192 & $1.22\times 10^{-4}$ & $3.92\times 10^{-12}$ \\
  $128$--$512$ & 4096 &  $2.44\times 10^{-4}$ & $7.83\times 10^{-12}$ \\
  $512$--$1024$ & 2048 &  $4.88\times 10^{-4}$ & $1.57\times 10^{-11}$ \\
  $1024$--$2048$ & 1024 &  $9.76\times 10^{-4}$ & $3.13\times 10^{-11}$ \\
 \hline
\end{tabular}
\caption[FH FFT]{Properties of the FFTs used in the \fh\ pipeline. The time duration $T_{\text{FFT}}$ refers to the length in seconds of the data chunks on which the FFT is computed. The frequency bin width is the inverse of the time duration, while the spin-down bin width is computed as $\delta \dot{f}={\delta
f}/{T_{\textrm{obs}}}$, where $T_{\textrm{obs}}$ is the total run duration.}
\label{tab:fh_fft}
\end{center}
\end{table}

The time-frequency peaks of the peakmap are properly shifted, for each sky position\footnote{Over a suitable grid, which bin size depends on the frequency and sky location.}, to compensate the Doppler effect due to the detector motion~\cite{2014PhRvD..90d2002A}. The shifted peaks are then fed to the \fh\ algorithm~\cite{2014PhRvD..90d2002A}, which transforms each peak to the frequency/spin-down plane of the source. The frequency and spin-down bins (which we will refer to as {\it coarse} bins in the following) depend on the frequency band, as shown in Table~\ref{tab:fh_fft}, and are defined, respectively, as $\delta f ={1}/{T_{\text{FFT}}}$ and  $\delta \dot{f}={\delta
f}/{T_{\textrm{obs}}}$, where $T_{\textrm{obs}}$ is the total run duration.
In practice, the nominal frequency resolution has been increased by a factor of 10 \cite{2014PhRvD..90d2002A}, as the {\fh} is not computationally bounded by the width of the frequency bin. The algorithm, moreover, adaptively weights any noise non-stationarity and the time-varying detector response~\cite{2005CQGra..22S1255P}.  

The whole analysis is split into tens of thousands of independent jobs, each of which covers a
small portion of the parameter space. Moreover, for frequencies above 512 Hz a GPU-optimized implementation of the {\fh} transform has been used~\cite{2021Univ....7..218L}.

The output of a \fh\, transform is a 2-D histogram in the frequency/spin-down plane of the source.

Outliers, that is significant points in this plane, are selected by dividing each 1 Hz band of the corresponding histogram into 20 intervals and taking, for each interval, and for each sky location, the one or (in most cases) two candidates with the highest histogram number count \cite{2014PhRvD..90d2002A}. All the steps described so far are applied separately to  the data of each detector involved in the analysis.

As in past analyses~\cite{2017PhRvD..96f2002A,2019PhRvD.100b4004A}, candidates from each detector are clustered and then coincident candidates among the clusters of a pair of detectors are found using a distance metric\footnote{The metric is defined as
$$d_{\rm FH}=\sqrt{\left(\frac{\Delta f}{\delta f}\right)^2+\left(\frac{\Delta \dot{f}}{\delta \dot{f}}\right)^2+\left(\frac{\Delta \lambda}{\delta \lambda}\right)^2+\left(\frac{\Delta 
\beta}{\delta \beta}\right)^2},$$ $\Delta f$, $\Delta \dot{f}$, $\Delta \lambda$, and $\Delta \beta$ are the differences, for each parameter, among pairs of candidates of the two detectors, and $\delta f$, $\delta 
\dot{f}$, $\delta \lambda$, and $\delta \beta$ are the corresponding bin widths.}
$d_{\rm FH}$ built in the four-dimensional
parameter space of sky position $(\lambda,~\beta)$ (in ecliptic coordinates), frequency $f$ and spin-down $\dot{f}$. Pairs of candidates with distance  $d_{\rm FH}\le 3$ are considered coincident. In the current O3 analysis, coincidences have been done only among the two LIGO detectors for frequencies above 128 Hz, while also coincidences H1 - Virgo and L1 - Virgo have been considered for frequencies below 128 Hz, where the difference in sensitivity (especially in the very low frequency band) is less pronounced.

Coincident candidates are ranked according to the value of a statistic built using the distance and the \fh\ histogram weighted number count of the coincident candidates \cite{2014PhRvD..90d2002A}. After the ranking, the eight outliers in each 0.1 Hz band with the highest values of the statistic are selected and subject to the follow-up. 
\subsubsection{Follow-up}
\label{subse:fh_fu}
The \fh\ follow-up runs on each outlier of each coincident pair. It is based on the construction of a new peakmap, over $\pm 3$ {\it coarse} bins around the frequency of the outlier, with a longer $T_{\rm FFT}$. This new peakmap is built after the removal of the signal frequency variation due to the Doppler effect for a source located at the outlier sky position.\\
A new refined grid on the sky is built around this point, covering $\pm 3$ {\it coarse} bins, in order to take into account the uncertainty on the outlier parameters. For each point of this grid we remove the residual Doppler shift from the peakmap by properly shifting the frequency peaks. Each new corrected peakmap is the input for the \fh\, transform to explore the frequency and the spin-down range of interest ($\pm 3$ {\it coarse} bins for the frequency and the spin-down). 
The most significant peak among all the \fh\, histograms, characterized by a set of {\it refined} parameters, is selected and subject to further post-processing steps.

First, the {\it significance} veto (V1) is applied. It consists in building a new peakmap over $0.2$ Hz around the outlier refined frequency, after correcting the data with its refined parameters. The corrected peakmap is then projected on the frequency axis. Its frequency range is divided in sub-bands, each covering $\pm 2$ coarse frequency bins. 
The maximum of the projection in the sub-band containing the outlier is compared with the maxima selected in the remaining {\it off-source} intervals. The outlier is kept if it ranks as first or second for both detectors.
\\Second, a {\it noise line} veto (V2) is used, which discards outliers whose frequency, after the removal of the Doppler and spin-down corrections, overlaps a band polluted by known instrumental disturbances.\\
The {\it consistency} test (V3) discards pairs of coincident outliers if their Critical Ratios (CRs), properly weighted by the detector noise level, differ by more than a factor of 5. The CR is defined as
\begin{equation}
CR = \frac{x-\mu}{\sigma},
\label{eq:crdef}
\end{equation}
where $x$ is the value of the peakmap projection in a given frequency bin, $\mu$ is the average value and $\sigma$ the standard deviation of the peakmap projection.
\\
The {\it distance} veto (V4) consists in removing pairs of coincident outliers with distance $d_{\rm FH}>6$ after the follow-up.\\
Finally, outliers with distance $d_{\rm FH}<3$ from {\it hardware injections} are also vetoed (V5).\\
Outliers which survive all these vetoes are scrutinized more deeply, by applying a further follow-up step, based on the same procedures just described, but further increasing the segment duration $T_{\rm FFT}$. 

\subsubsection{Parameter space}
The \fh\ search covers the frequency range [10, 2048] Hz, a spin-down range between -$10^{-8}$ Hz/s to $10^{-9}$ Hz/s and the whole sky. The frequency and spin-down resolutions are given in Tab. \ref{tab:fh_fft}. The sky resolution, on the other hand, is a function of the frequency and of the sky position and is defined in such a way that for two nearby sky cells the maximum frequency variation, due to the Doppler effect, is within one frequency bin, see \cite{2014PhRvD..90d2002A} for more details.
\subsubsection{Upper limits}\label{sec:fhulmethod}
Upper limits are computed for every 1~Hz sub-band in the range of 20--2048~Hz\footnote{Although the search starts at 10 Hz, we decided to compute upper limit starting from 20 Hz, due to the unreliable calibration at lower frequency.}, considering only the LIGO detectors, as Virgo sensitivity is worse for most of the analyzed frequency band. First, for each detector we use the analytical relation \cite{2014PhRvD..90d2002A}
\be
h_{UL,95\%}\approx
\frac{4.97}{N^{1/4}}\sqrt{\frac{S_n(f)}{T_{\rm FFT}}}\sqrt{CR_{\rm max}+1.6449},
\label{eq:hul}
\ee 
where $N=T_{\rm obs}/T_{\rm FFT}$, $S_n(f)$ is the detector average noise power spectrum and $CR_{\rm max}$ is the maximum outlier CR\footnote{Defined by Eq. \ref{eq:crdef} and where in this case the various quantities are computed over the Frequency-Hough map}, in the given 1 Hz band. For each 1 Hz band, the final upper limit is the worse among those computed separately for Hanford and Livingston.

As verified through a detailed comparison based on LIGO and Virgo O2 and O3 data, this procedure produces conservative upper limits with respect to those obtained through the injection of simulated signals, which is computationally much heavier \cite{DicesareThesis}.

Moreover, it has been shown that the upper limits obtained through injections are always above those based on Eq. \ref{eq:hul} when the minimum CR in each 1 Hz sub-band is used. The two curves based, respectively, on the highest and the smallest CR delimit a region containing both a more stringent upper limit estimate and the search sensitivity estimate, that is the  minimum strain of a detectable signal. Any astrophysical implication of our results, discussed in Sec. \ref{sect:results} will be always based on the most conservative estimate.

\subsection{\sh}
\label{ssect:SHm}

\sh{} \cite{2004PhRvD..70h2001K, Sintes:2006uc} is a semicoherent pipeline based on the Hough transform to look for CW signals from isolated neutron stars. 
Several versions of this pipeline have been used throughout the initial \cite{2008PhRvD..77b2001A,2016PhRvD..94d2002A}
and advanced \cite{2017PhRvD..96f2002A, 2018PhRvD..97j2003A} detector era,
as well as to look for different kinds of signals such as CW from neutron stars in binary systems \cite{Covas:2019jqa, Covas:2020nwy, LIGOScientific:2020qhb}
or long-duration GW transients \cite{Oliver:2019ksl}.
The current implementation of \sh{} closely follows that of \cite{2018PhRvD..97j2003A} and includes an improved suite of post-processing and follow-up stages 
\cite{Tenorio:2020cqm,2021JOSS....6.3000K,Tenorio:2021njf}.

\subsubsection{Parameter space}

The \sh{} pipeline searches over the standard four parameters describing a CW signal from isolated
NS: frequency $f$, spin-down $\dot{f}$ and sky position, 
parametrized using equatorial coordinates $\alpha, \delta$.

Parameter-space resolutions are given in \cite{2004PhRvD..70h2001K}
\begin{equation}
    \delta f = \frac{1}{\Tsft} ,\quad 
    \delta \dot{f} = \frac{\delta f}{\Tobs} ,\quad
    \delta \theta = \frac{c/v}{\Tsft \; \pf \; f} ,\quad
    \label{eq:sh_resolution}
\end{equation}
where $\theta$ represents either of the sky angles, 
$v/c \simeq 10^{-4}$ represents the average detector velocity as a fraction of the speed of light,
and the pixel factor $\pf = 2$ is a tunable overresolution parameter.  
Table~\ref{tab:sh_resolution} summarizes the numerical values employed in this search.

\begin{table}[]
    \centering
    \begin{tabular}{cl}
        \toprule
            Parameter  & Resolution \\
        \midrule
            $\delta f$ & $1.4 \times 10^{-4}$ Hz\\
            $\delta \dot{f}$ & $5 \times 10^{-12}$ Hz/s\\
            $\delta \theta$ & $0.69\;\textrm{Hz} / f$\\
        \bottomrule
    \end{tabular}
    \caption{Parameter-space resolutions employed by the \sh{} pipeline.}
    \label{tab:sh_resolution}
\end{table}

The \sh{} all-sky search covers the most sensitive frequency band of the advanced LIGO detectors, 
between 65 Hz and 350 Hz. This band is further sub-divided into \mbox{$\Delta f = 0.025\;\textrm{Hz}$} sub-bands,
resulting in a total of 11400 frequency bands.
Spin-down values are covered from $-1 \times 10^{-9}\; \textrm{Hz} / \textrm{s}$ to $5 \times 10^{-12}\; \textrm{Hz} / \textrm{s}$,
which include typical spin-up values associated to CW emission from the evaporation of boson clouds
around black holes \cite{Zhu:2020tht}.

\subsubsection{Description of the search}

The first stage of the \sh{} pipeline performs a multi-detector search using H1 and L1 SFTs with
\mbox{$\Tsft=7200s$}. Each $0.025\;\textrm{Hz}$ sub-band is analyzed separately using the same two step strategy as in \cite{2018PhRvD..97j2003A, LIGOScientific:2020qhb}: parameter-space is efficiently analyzed using \sh{}'s look-up table approach;
the top 0.1\% most significant candidates are further analyzed using a more sensitive statistic. 
The result for each frequency sub-band is a toplist containing the $10^5$ most significant candidates across the sky and spin-down parameter-space. 

Each toplist is then clustered using a novel approach presented in \cite{Tenorio:2020cqm} and firstly
applied in \cite{LIGOScientific:2020qhb}. A parameter-space distance is defined using the average mismatch in frequency 
evolution between two different parameter-space templates
\begin{equation}
    d(\vec{\lambda}, \vec{\lambda}_{*}) = \frac{T_{\textrm{SFT}}}{N_{\textrm{SFT}}} \sum_{\alpha=0}^{N_{\textrm{SFT}}}\left|f(t_{\alpha};\vec{\lambda}) - f(t_{\alpha};\vec{\lambda}_{*}) \right| \;,
\end{equation}
where $f(t;\vec{\lambda})$ is defined as
\begin{equation}
f(t;\vec{\lambda}) 
= \left[
f + (t - t_{\textrm{ref}}) \cdot \dot{f}
\right] \cdot \left[
1 + \frac{\vec{v}(t) \cdot \vec{n}}{c} 
\right]\;
\end{equation}
and $\vec{\lambda}=\{f, \dot{f}, \alpha, \delta\}$ refers to the phase-evolution 
parameters of the template.

Clusters are constructed by pairing together templates in consecutive frequency bins such that $d(\vec{\lambda}, \vec{\lambda}_{*}) \leq 1$.
Each cluster is characterized by its most significant element  (the \emph{loudest} element). 
From each $0.025\;\textrm{Hz}$ sub-band, we retrieve the forty most significant clusters for further analysis. 
This results in a total of $456000$ candidates to follow-up.

The loudest cluster elements are first sieved through the \emph{line veto}, a standard tool to discard clear
instrumental artifacts using the list of known, narrow, 
instrumental artifacts (lines) in the advanced LIGO detectors \cite{bib:O3Linelist}: 
If the instantaneous frequency of a candidate overlaps with a frequency band containing an instrumental line of known origin, 
the candidate is ascribed an instrumental origin and consequently ruled out. 

Surviving candidates are then followed-up using \texttt{PyFstat}, 
a Python package implementing a Markov-chain Monte Carlo (MCMC) 
search for CW signals \cite{2018PhRvD..97j3020A, 2021JOSS....6.3000K}. 
The follow-up uses the \mbox{$\mathcal{F}$-statistic} as a (log) Bayes factor to sample the posterior probability distribution of the phase-evolution parameters
around a certain parameter-space region
\begin{equation}
    \mathrm{P}(\vec{\lambda}|x) \propto e^{\mathcal{F}(\vec{\lambda}; x)} \cdot \mathrm{P}(\vec{\lambda}) \;,
\end{equation}
where $\mathrm{P}(\vec{\lambda})$ represents the prior probability distribution of the phase-evolution parameters. The~\mbox{$\mathcal{F}$-statistic}, 
as opposed to the \sh{} number count, 
allows us to use longer coherence times, 
increasing the sensitivity of the follow-up with respect to the main search stage.

As initially described in \cite{2018PhRvD..97j3020A}, the effectiveness of an MCMC follow-up
is tied to the number of templates covered by the initial prior volume, 
suggesting a hierarchical approach: coherence time should be increased following a ladder
so that the follow-up is able to converge to the true signal parameters at each stage.
We follow the proposal in~\cite{Tenorio:2021njf} and compute a coherence-time ladder using $\mathcal{N}^{*} = 10^{3}$ (see Eq. (31) of \cite{2018PhRvD..97j3020A}) starting from  $\Tcoh = 1\;\textrm{day}$ 
including an initial stage of $\Tcoh = 0.5\;\textrm{days}$.
The resulting configuration is collected in Table \ref{tab:sh_t_coh}.

\begin{table}
    \centering
    \begin{tabular}{ccccccc}
        \toprule
        Stage & 0 & 1 & 2 & 3 & 4 & 5 \\
        \midrule
        $N_{\textrm{seg}}$ & 660 & 330 & 92 & 24 & 4 & 1 \\
        \midrule
        $\Tcoh$ & 0.5 day & 1 day & 4 days & 15 days & 90 days & 360 days \\
        \bottomrule
    \end{tabular}
    \caption{Coherence-time configuration of the multi-stage follow-up
    employed by the \sh{} pipeline.
    The data stream is divided into a fix number of segments of the same length;
    the reported coherence time is an approximate value obtained by dividing the 
    observation time by the number of segments at each stage. }
    \label{tab:sh_t_coh}
\end{table}

The first follow-up stage is similar to that employed 
in~\cite{Covas:2020nwy, LIGOScientific:2020qhb}: 
an MCMC search around the loudest candidate of the selected clusters is performed using a 
coherence time of $\Tcoh=0.5$ days. Uniform priors containing 4 parameter-space bins in each 
dimension are centered around the loudest candidate.
A threshold is calibrated using an injection campaign: any candidate whose loudest
$2\mathcal{F}$ value over the MCMC run is lower than $2\mathcal{F} = 3450$ 
is deemed inconsistent with CW signal.

The second follow-up stage is a variation of the method described in~\cite{Tenorio:2021njf},
previously applied to \cite{LIGOScientific:2021tsm, LIGOScientific:2021ozr}.
For each outlier surviving the initial follow-up stage 
(stage 0 in Table~\ref{tab:sh_t_coh}), we construct a Gaussian prior using the median and
inter-quartile range of the posterior samples and run the next-stage MCMC follow-up.
The resulting maximum $2\mathcal{F}$ is then compared to the expected $2\mathcal{F}$ 
inferred from the previous MCMC follow-up stage. 
Highly-discrepant candidates are deemed inconsistent with a CW signal and hence discarded.

Given an MCMC stage using $\hat{N}$ segments from which a value of 
$2\hat{\mathcal{F}}$ is recovered, the distribution of $2\mathcal{F}$ values using 
$N$ segments is well approximated by
\begin{equation}
    \mathrm{P}(2\mathcal{F}|N, 2\hat{\mathcal{F}}, \hat{N}) 
    = \textrm{Gauss}(2\mathcal{F} ; \mu, \sigma) \;,
    \label{eq:sh_P_S}
\end{equation}
where
\begin{equation}
    \mu = \rho_0^2 + 4 N \;,
\end{equation}
\begin{equation}
    \sigma^2 =  8 \cdot (N + \hat{N} + \rho_0^2) \;,
\end{equation}
and $\rho_0^2 = 2\hat{\mathcal{F}} - 4 \hat{N}$
is a proxy for the (squared) SNR~\cite{prix_coherent_semicoherent}.
Equation~\eqref{eq:sh_P_S} is exact in the limit of \mbox{$N, \hat{N} \gg 1$} 
or \mbox{$\rho_0^2 \gg 1$}.
In this search, however, we calibrate a bracket on $(2\mathcal{F} - \mu)/\sigma$ for each
follow-up stage using an injection campaign, shown in table \ref{table:sh_fu_brackets}.
Candidates outside of the bracket are deemed inconsistent with a CW signal.

\begin{table}
    \centering
    \begin{tabular}{cc}
        \toprule
        Comparing stages & $(2\mathcal{F} - \mu)/\sigma$ bracket \\
        \midrule
        Stage 0 v.s. Stage 1 & (-1.79, 1.69)\\
        Stage 1 v.s. Stage 2 & (-1.47, 1.35)\\
        Stage 2 v.s. Stage 3 & (-0.94, 0.80)\\
        Stage 3 v.s. Stage 4 & (-0.63, 0.42)\\
        Stage 4 v.s. Stage 5 & (-0.34, 0.11)\\
        \bottomrule
    \end{tabular}
    \caption{$2\mathcal{F}$ consistency brackets employed in the multi-stage follow-up
    of the \sh{} pipeline. Brackets were computed using a campaign of 500 software-injected
    signals representing an isotropic population of uniformly sky-distributed NS at 150
    representative frequency bands with an amplitude corresponding to the $h_0^{95\%}$
    sensitivity estimation. The implied false dismissal probability is 
    \mbox{$\lesssim 1 / (150 \times 500) \simeq 1.3 \times 10^{-5} $}. 
    Stages correspond to those described in Table~\ref{tab:sh_t_coh}.}
    \label{table:sh_fu_brackets}
\end{table}
Any surviving candidates are subject to manual inspection in search for
obvious instrumental causes such as hardware-injected artificial signals
or narrow instrumental artifacts.

\subsection{\tdfstat}
\label{ssect:FStatm}

The {\tdfstat} search method has been applied to an all-sky search of VSR1 data~\cite{Aasi:2014} and all-sky searches of the LIGO O1 and O2 data~\cite{2017PhRvD..96f2002A,2018PhRvD..97j2003A,2019PhRvD.100b4004A}.  
The main tool of the pipeline is the $\F$-statistic \cite{1998PhRvD..58f3001J} with which one can coherently search the data over a reduced parameter space consisting of signal frequency, its derivatives, and the sky position of the source. 
However, a coherent all-sky search over the long data set like the whole data of O3 run is computationally prohibitive.
Thus the data are divided into shorter time domain segments.
Moreover, to reduce the computer memory required to do the search, the data are divided into narrow-band segments that are analyzed separately.
As a result the {\tdfstat} pipeline consists of two parts. The first
part is the coherent search of narrowband, time-domain segments. The second part is the search for coincidences among the parameters of the candidates obtained from the coherent search of all the time domain segments.

The algorithms to calculate the $\F$-statistic in the coherent search are described in Sec.\ 6.2 of~\cite{Aasi:2014}.
The time series is divided into segments, called frames, of six sidereal days long each. 
Moreover the data are divided into sub-bands of 0.25 Hz overlapped by 0.025 Hz.
The O3 data has a number of non-science data segments. The values of these bad data are set to zero. For our analysis, we choose only segments that have a fraction of bad data less than 60\% both in H1 and L1 data
and there is an overlap of more than 50\% between the data in the two detectors. This requirement results in forty-one 6-day-long data segments for each sub-band. 
For the search we use a four-dimensional grid of templates (parameterized by frequency, spin down rate, and two more parameters related to the position of the source in the sky) constructed in Sec.\ 4 of~\cite{pisarski:2015} with grid's minimal match parameter $\textrm{MM}$ chosen to be $\sqrt{3}/2$. 
This choice of the grid spacing led to the following resolution for the four parameters of the space that we  search
\begin{subequations}
\begin{align}
  \Delta f &\simeq  1.9\times 10^{-6}~\text{Hz},\\
  \Delta \dot{f} &\simeq 1.1\times 10^{-11}~\text{Hz/s}, \\
  \Delta \alpha &\simeq 7.4\times 10^{-2} \left(\frac{100~\text{Hz}}{f}\right) \text{rad},\\
  \Delta \delta &\simeq 1.5\times 10^{-2} \left(\frac{100~\text{Hz}}{f}\right) \text{rad}.
\end{align}
\end{subequations}
  
We set a fixed threshold of 15.5 for the $\F$-statistic and record the parameters of all threshold crossings,
together with the corresponding values of the $\F$-statistic.
In the second stage of the analysis we use exactly the same coincidence search algorithm as in the analysis of VSR1 data
and described in detail in Sec.\ 8 of~\cite{Aasi:2014} with only one change.
We use a different coincidence cell from that described in~\cite{Aasi:2014}.
In~\cite{Aasi:2014} the coincidence cell was constructed from Taylor expansion of the autocorrelation function of the $\F$-statistic.
In the search performed here the chosen coincidence cell is a suitably scaled grid cell used in the coherent part of the pipeline.
We scale the four dimensions of the grid cell by different factors given by [16 8 2 2] corresponding to
frequency, spin down rate (frequency derivative), and two more parameters related to the position of the source in the sky respectively.
This choice of scaling gives optimal sensitivity of the search.  
We search for coincidences in each of the bands analyzed. 
Before identifying coincidences we veto candidate signals overlapping with the instrumental lines identified by independent analysis of the detector data. 
To estimate the significance of a given coincidence, we use the formula for the false alarm probability derived in the appendix of~\cite{Aasi:2014}.
Sufficiently significant coincidences are called outliers and are subject to a further investigation.

\subsubsection{Parameter space}

Our \tdfstat \, analysis is a search over a 4-dimensional space consisting of four parameters: frequency, spin-down rate and sky position.
As we search over the whole sky the search is very computationally intensive. Given that our computing resources are limited, to achieve a satisfactory sensitivity we have restricted the range of frequency and spin-down rates analyzed to cover the frequency and spin-down ranges of the bulk of the observed pulsars. Thus we have searched the gravitational frequency band from 20~Hz to 750~Hz. The lower frequency of 20 Hz is chosen due to the low sensitivity of the interferometers below 20~Hz. In the frequency 20~Hz to 130~Hz range, assuming that the GW frequency is twice the spin frequency, we cover young and energetic pulsars, such as Crab and Vela. In the frequency range from 80 Hz to 160~Hz we can expect GW signal due to r-mode instabilities \citep{Owen1998,Alford14}. In the frequency range from 160~Hz to 750~Hz we can expect signals from most of the recycled millisecond pulsars, see Fig.\ 3 of \citep{Patruno_2017}.

For the GW frequency derivative $\dot{f}$ we have chosen a frequency dependent range.
Namely, for frequencies less than 200~Hz we have chosen $\dot{f}$ to be in the range $[-f/\tau_{min},0]$,
where $\tau_{min}$ is a limit on pulsar's characteristic age, and we have taken $\tau_{min}=1000$~yr.
For frequencies greater than 200~Hz we have chosen a fixed range for the spin-down rate.
As a result, the following ranges of $\dot{f}$ were searched in our analysis:
\begin{subequations}
\begin{align}
0 > \,&\dot{f}\, > - 3.2\times 10^{-9}\frac{f}{100~\text{Hz}} \text{Hz/s},
\nonumber\\[1ex]
&\qquad\qquad \text{for}\ f < 200~\text{Hz},
\\[1ex]
2{\times}10^{-11}~\text{Hz/s} > \,&\dot{f}\, > -2{\times}10^{-10}~\text{Hz/s},
\nonumber\\
&\qquad\qquad \text{for}\ f > 200~\text{Hz}.
\end{align}
\end{subequations}

\begin{figure} 
\includegraphics[width=\columnwidth]{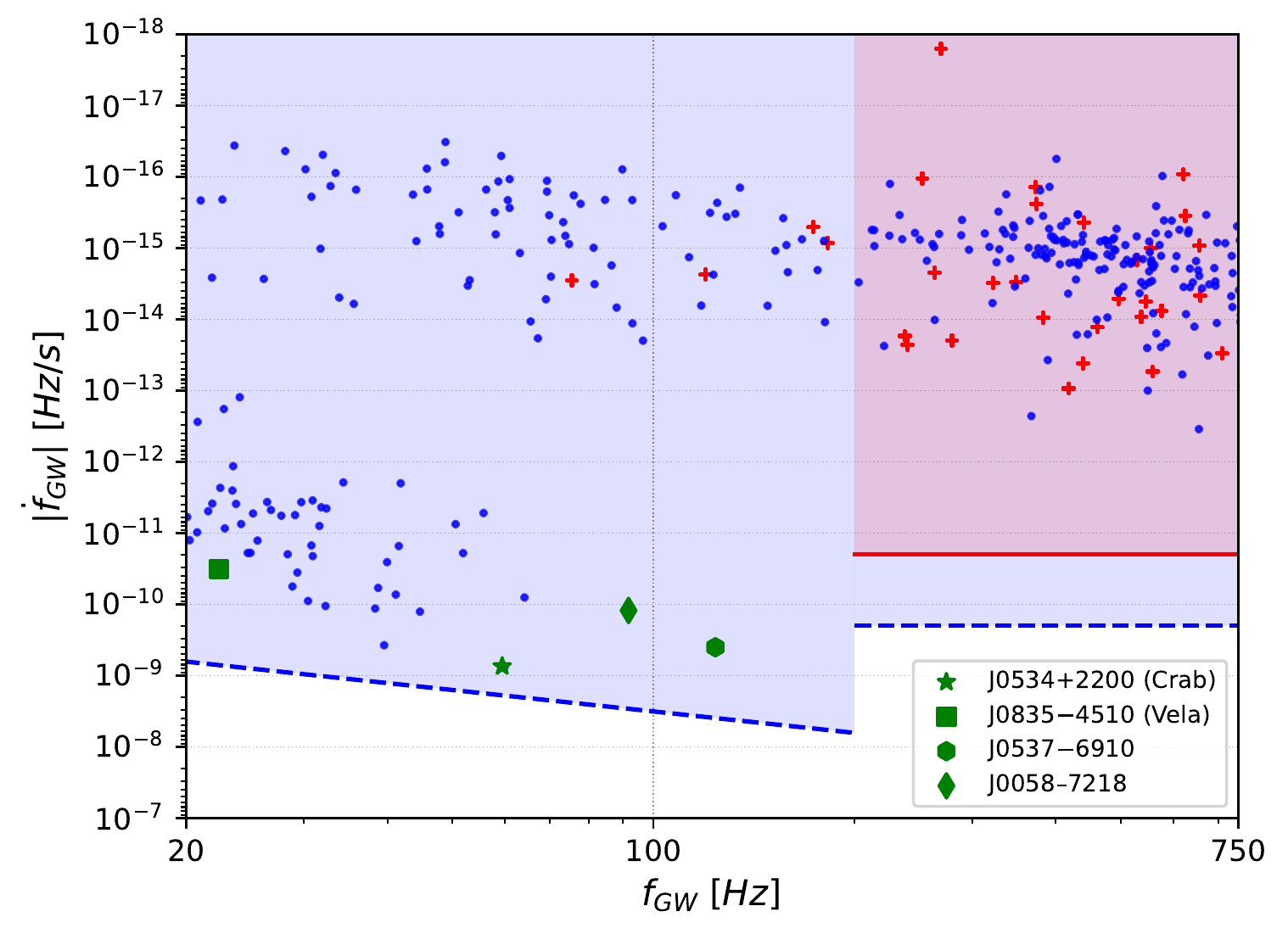}
\caption{Frequency time derivative for tentative emission of GWs ($\dot{f} \equiv 2\dot{f}_{rot}$) as a function of the frequency of emitted GWs ($f \equiv 2f_{rot}$), where $f_{rot}$ and $\dot{f}_{rot}$ are rotational frequency and frequency time derivative for known pulsars, obtained from the Australia Telescope National Facility (ATNF) database \cite{2005AJ....129.1993M}. The vertical axis shows the absolute value for both negative values of the frequency time derivative (``spin-down'', blue dots) and positive values (``spinup'', red plus symbols). Blue dashed lines represent spin-down limits used in the {\tdfstat} search: for $f < 200$ Hz, $0 > \dot{f} > -f/\tau_{min}$, where $\tau_{min} = 1000$~yr denotes a limit on pulsar's characteristic age; for $f > 200$~Hz, $\dot{f} > -2{\times}10^{-10}$~Hz/s. For $f > 200$~Hz in the case of spinning-up objects, in the $\F$-statistic search we admit a positive range of values to $\dot{f} < 2{\times}10^{-11}$~Hz/s. The boundary of this range is marked by a red continuous line. } 
\label{fig:ffdot_gw} 
\end{figure}

In Fig.~\ref{fig:ffdot_gw} we plot GW frequency derivatives against GW frequencies (assuming the GW frequency is twice the spin frequency of the pulsar) for the observed pulsars from the ATNF catalogue \cite{2005AJ....129.1993M}.
We show the range of the GW frequency derivative selected in our search, and one can see that the expected frequency derivatives of the observed pulsars are well within this range. Note, finally, that we have made the conservative choice of including positive values of the frequency derivative (`spin-up'), in order to search as wide a range as possible. In most cases, however, the pulsars that appear to spin-up are in globular clusters, for which the local forces make the measurement unreliable \cite{Freire2017}.

\subsubsection{Sensitivity of the search} 
\label{sssec:Fstatsen}

\begin{figure}
\centering
\includegraphics[width=\columnwidth]{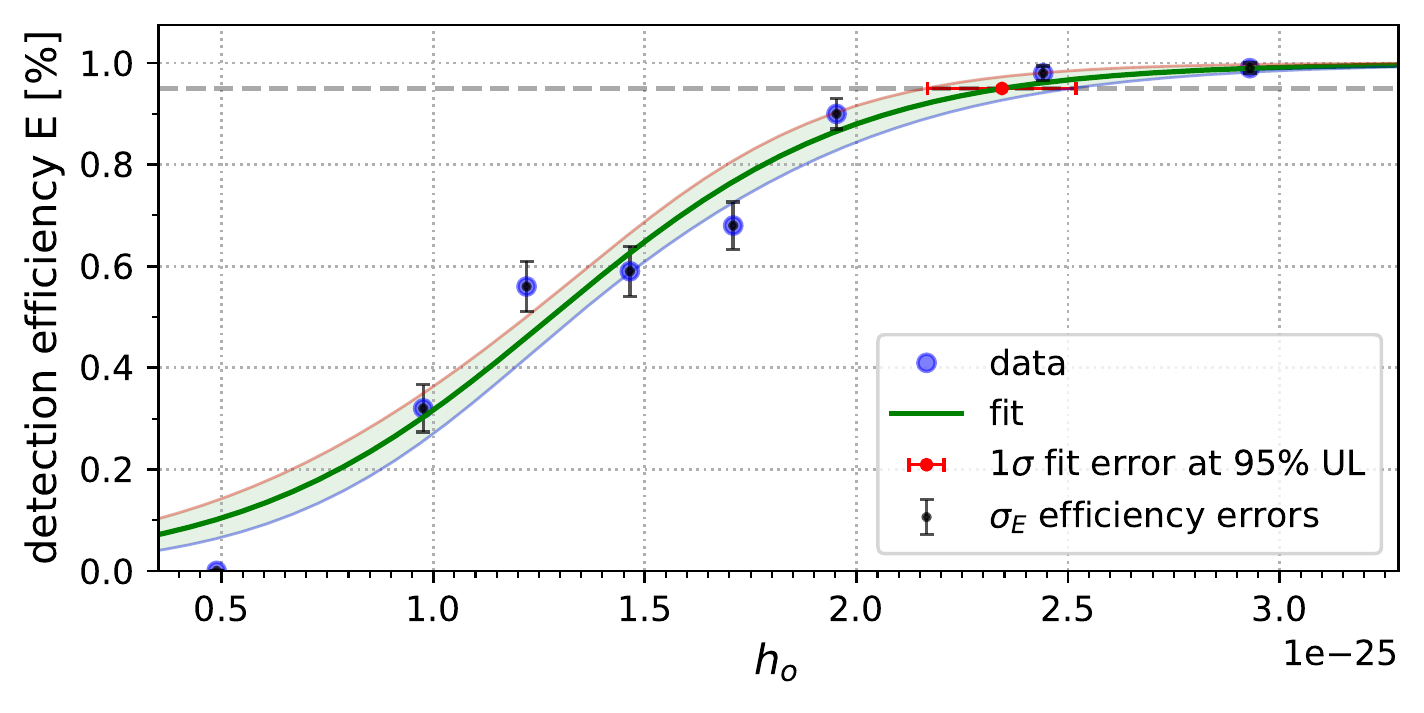}
\caption{Example sigmoid function fit (green solid line) to the injected data efficiencies (blue dots),
representing the detection efficiency $E$ as a function of injected GW amplitude $h_0$ used in {\tdfstat} search.
Pale red and blue curves mark the $1\sigma$ confidence band obtained from the uncertainty of the fit.
Red error bar marks the ${\pm}1\sigma$ standard deviation on the $h_0^{95\%}$ value, corresponding to the efficiency of 0.95 (indicated by the horizontal dashed gray line).
Vertical errors for each efficiency represent $1\sigma$ standard binomial errors related to detection rate, $\sigma_E = \sqrt{E(1-E)/N_i}$,
where $E$ is the efficiency and $N_i=100$ is the number of injections for each GW amplitude.
The data shown relates to the sub-band with frequency of the lower edge of the band equal to 725.95~Hz.}
\label{fig:tdfstat_ul_sigmoid_example}
\end{figure}

In order to assess the sensitivity of the {\fstat} search, we set upper limits on the intrinsic GW amplitude $h_0$ in each 0.25~Hz bands. To do so, we generate signals for an array of 8 amplitudes $h_0$ and for randomly selected sky positions (samples drawn uniformly from the sphere). For each amplitude, we generate 100 signals with $f$, $\dot{f}$, the polarization angle $\psi$ and cosine of the inclination angle $\iota$ are chosen from uniform random distributions in their respective ranges. The signals are added to the real data segments, and searches are performed with the same grids and search set-up as for the real data search, in the neighbourhood of injected signal parameters. We search $\pm 6$ grid points for $\dot{f}$ and $\pm 1$ grid points for the sky positions away from the true values of the signal's parameters. 
We consider a signal {\bf detected} if {\em coincidence multiplicity for the injected signal is higher than the highest signal multiplicity in a given sub-band and in a given hemisphere in the real data search}.
The detection efficiency is the fraction of recovered signals. We estimate the $h_0^{95\%}$, i.e., 95\% confidence upper limit on the GW amplitude $h_0$, by fitting\footnote{For the $h_0^{95\%}$ fitting procedure, we use the {\tt python 3} \citep{10.5555/1593511} {\tt scipy-optimize} \citep{2020SciPy-NMeth} {\tt curve\_fit} package, implementing the Levenberg-Marquardt least squares algorithm, to obtain the best fitted parameters, $x_0$ and $k$, to the sigmoid function. Errors of parameters $\delta x_0$ and $\delta k$ are obtained from the covariance matrix and used to calculate the standard deviation $\sigma_e$ of the detection efficiency as a function of $h_0$ i.e., the confidence bands around the central values of the fit. In practice, we use the {\tt uncertainties} package \citep{uncertainties} to obtain the ${\pm}1\sigma$ standard deviation on the $h_0$ value.} a sigmoid function to a range of detection efficiencies $E$ as a function of injected amplitudes $h_0$, $E(h_0) = \left(1 + e^{k(x_0 - h_0)}\right)^{-1}$, with $k$ and $x_0$ being the parameters of the fit. Figure~\ref{fig:tdfstat_ul_sigmoid_example} presents an example fit to the simulated data with $1\sigma$ errors on the $h_0^{95\%}$ estimate marked in red.

\subsection{\soap}
\label{ssect:SOAPm}

{\soap} \cite{2019PhRvD.100b3006B} is a fast, model-agnostic search for long duration signals based on the Viterbi algorithm \cite{Viterbi1967}. 
It is intended as both a rapid initial search for isolated NSs, quickly providing candidates for other search methods to investigate further, as well as a method to identify long duration signals which may not follow the standard \ac{CW} frequency evolution.
In its most simple form {\soap} analyzes a spectrogram to find the continuous time-frequency track which gives the highest sum of fast Fourier transform power.
If there is a signal present within the data then this track is the most likely to correspond to that signal.
The search pipeline consists of three main stages, the initial {\soap} search \cite{2019PhRvD.100b3006B}, the post processing step using convolutional neural networks \cite{Bayley2020} and a parameter estimation stage. 

\subsubsection{Data preparation \label{ssect:SOAPm:data}}
The data used for this search starts as calibrated detector data which is used to create a set of \acp{FFT} with a coherence time of 1800 s. 
The power spectrum of these \acp{FFT} are then summed over one day, i.e. every 48 \acp{FFT}. 
Assuming that the signal remains within a single bin over the day, this averages out the antenna pattern modulation and increases the SNR in a given frequency bin.
As the frequency of a \ac{CW} signal increases, the magnitude of the daily Doppler modulation also increases, therefore the assumption that a signal remains in a single frequency bin within one day no longer holds. 
Therefore, the analysis is split into 4 separate bands (40-500 Hz, 500-1000 Hz, 1000-1500 Hz, 1500-2000 Hz) where for each band the Doppler modulations are accounted for by taking the sum of the power in adjacent frequency bins.
For the bands starting at 40, 500, 1000 and 1500 Hz, the sum is taken over every one (no change), two, three and four adjacent bins respectively such that the resulting time-frequency plane has one, two, three or four times the width of bin.
The data is then split further into `sub-bands' of widths 0.1, 0.2, 0.3 and 0.4 Hz wide respective to the four band sizes above. 
These increase in width such that the maximum yearly Doppler shift $\Delta f_{\rm{orb}}^{(\rm{max})}$ is half the sub-band width, where the maximum is given by
\begin{equation}
    \Delta f_{\rm{orb}}^{(\rm{max})} = f\frac{v_{\rm{orb}}^{\rm{(max)}}}{c} \approx 10^{-4} f,
\end{equation}
where $v_{\rm{orb}}^{\rm{(max)}}$ is the maximum orbital velocity of the earth relative to the source, $c$ is the speed of light and $f$ is the initial pulsar frequency.
Each of the sub-bands are overlapping by half of the sub-band width such that any signal should be fully contained within a sub-band.

\subsubsection{Search pipeline \label{ssect:SOAPm:SOAP}}

{\soap} searches through each of the summed and narrow-banded spectrograms described in Sec.~\ref{ssect:SOAPm:data} by rapidly identifying the track through the time frequency plane which gives the maximum sum of some statistic.
In this search the statistic used is known as the `line aware' statistic~\cite{2019PhRvD.100b3006B}, which uses multiple detectors data to compute the Bayesian statistic $p(\rm{signal})/[p(\rm{noise}) + p(\rm{line})]$, penalising instrumental line-like combinations of spectrogram powers. 
Since each of the four bands described in Sec.~\ref{ssect:SOAPm:data} take the sum of a different number of \ac{FFT} bins, the $\chi^2$ distributions that make up the Bayesian statistic are adjusted such that they have $2\times N\times M$ degrees of freedom, where $M$ is the number of summed frequency bins and $N$ is the number of summed time segments.

{\soap} then returns three main outputs for each sub-band: the Viterbi track, the Viterbi statistic and a Viterbi map. 
The Viterbi track is the time-frequency track which gives the maximum sum of statistics along the track, and is used for the parameter estimation stage in Sec.~\ref{ssect:SOAPm:parest}. 
The Viterbi statistic is the sum of the individual statistics along the track, and is one of the measures used to determine the candidates for followup in Sec.~\ref{ssect:SOAPm:cand}.
The Viterbi map is a time-frequency map of the statistics in every time-frequency bin which has been normalised along every time step. 
This is representative of the probability distribution of the signal frequency conditional on the time step and is used as input to the convolutional networks described in Sec.~\ref{ssect:SOAPm:CNN}.

\subsubsection{Convolutional neural network post processing\label{ssect:SOAPm:CNN}}

One post processing step in {\soap} consists of \acp{CNN} which take in combinations of three data types: the Viterbi map, the two detectors spectrograms and the Viterbi statistic.
The aim of this technique is to improve the sensitivity to isolated neutron stars by reducing the impact of instrumental artefacts on the detection statistic.
This part of the analysis does add some model dependency, so is limited to search for signals that follow the standard CW frequency evolution.
The structure of the networks are described in~\cite{Bayley2020}, where the output is a detection statistic which lies between 0 and 1.
These are trained on $\sim 1 \times 10^5$ examples of continuous wave signals injected into real data, where the data is split in the same way as described in Sec.~\ref{ssect:SOAPm:data}.
Each of the sub-bands is duplicated and a simulated continuous GW is injected into one of the two sub-bands such that the network has an example of noise and noise + signal cases. 
The sky positions, the frequency, frequency derivative, polarisation, cosine of the inclination angle and SNR of the injected signals are all uniformly drawn in the ranges described in~\cite{Bayley2020}.
These signals are then injected into real O3 data before the data processing steps described in Sec.~\ref{ssect:SOAPm:data}.
As the neural network should not be trained and tested on the same data, each of the training sub-bands are split into two categories (`odd' and `even'), where the sub-bands are placed in these categories alternately such that an 'odd' sub-band is adjacent to two `even' sub-bands. 
This allows a network to be trained on `odd' sub-bands and tested on `even' sub-bands and vice-versa.
The outputs from each of these networks can be combined and used as another detection statistic to be further analysed as described in Sec.~\ref{ssect:SOAPm:cand}.

\subsubsection{Candidate selection\label{ssect:SOAPm:cand}}

At this stage there is a set of Viterbi statistics and \ac{CNN} statistics for each sub-band that is analysed, from which a set of candidate signals need to be selected for followup.
Before doing this, any sub-bands which contain known instrumental artefacts are removed from the analysis.
The sub-bands corresponding to the top 1\% of the Viterbi statistics from each of the four analysis bands are then combined with the sub-bands corresponding to the top 1\% of CNN statistics, leaving us with a maximum of 2\% of the sub-bands as candidates.
It is at this point where we begin to reject candidates by manually removing sub-bands which contain clear instrumental artefacts and still crossed the detection threshold for either the Viterbi or CNN statistic.
There are a number of features we use to reject candidates including: strong detector artefacts which only appear in a single detectors spectrogram, broad ( $> 1/5$ sub-band width) long duration signals, individual time-frequency bins which contribute large amounts to the final statistic and very high power signals in both detectors. Examples of these features can be seen in section 6.3 of \cite{BayleyThesis2020}.
Any remaining candidates are then passed on for parameter estimation.

\subsubsection{Parameter estimation\label{ssect:SOAPm:parest}}

The parameter estimation stage uses the Viterbi track to estimate the Doppler parameters of the potential source. 
Due to the complicated and correlated noise which appears in the Viterbi tracks, defining a likelihood is challenging. 
To avoid this difficulty, likelihood-free methods are used, in particular a machine learning method known as a conditional variational auto-encoder.
This technique was originally developed for parameter estimation of compact binary coalescence signals \cite{gabbard2020bayesian}, and can return Bayesian posteriors rapidly ($< 1$s).
In our implementation, the conditional variational auto-encoder is trained on isolated NS signals injected into many sub-bands, and returns an estimate of the Bayesian posterior in the frequency, frequency derivative and sky position \cite{Bayley2021parest}.
This acts both as a further check that the track is consistent with that of an isolated NS, and provides a smaller parameter space for a followup search.

\section{Results}
\label{sect:results}
In this section we summarize the results of the search obtained by the four pipelines.
Each pipelines presents candidates obtained during the analysis and the results of the follow-up of the promising candidates. The upper limits on the GW strain are determined for each of the search procedures. 
There is also a study of the hardware injections of continuous wave signals added to the data.
During the O3 run 18 hardware injections were added to the LIGO data. The injections are denoted by ipN where N
is the consecutive number of the injection. The amplitudes of the injections added in the O3 run were significantly lower than those added in previous observing runs. Consequently the injections were more difficult to detect.

\subsection{\fh}
\label{ssect:FHr}
Outliers produced by the \fh\ search are followed-up with the procedure described in Sec. \ref{subse:fh_fu}. The increase in FFT duration sets the sensitivity gain of the follow-up step and it is mainly limited by the resulting computational load, which increases with the fourth power of $T_{\rm FFT}$ for a fixed follow-up volume.  
Moreover, $T_{\rm FFT}$ cannot be longer than about one sidereal day, because the current procedure is not able to properly deal with the sidereal splitting of the signal power, which would cause a sensitivity loss. 

All the coincident outliers produced by the \fh\ transform stage in the first frequency band, 10-128 Hz, have been followed-up. On the remaining frequency bands, from 128 Hz up to 2048 Hz, only outliers with $CR \geq 5$ (computed over the \fh\ map) in both detectors have been followed-up. This selection was also applied for pairs of coincident outliers produced in the L1 - Virgo and H1 - Virgo detectors in the frequency band 10-128 Hz.

\begin{figure*}
%\centering
\includegraphics[width=\columnwidth]{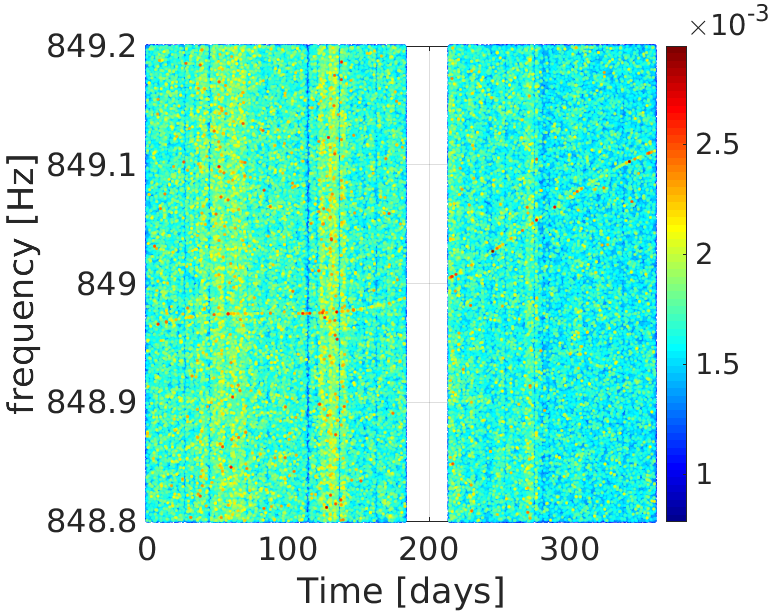}
\includegraphics[width=\columnwidth]{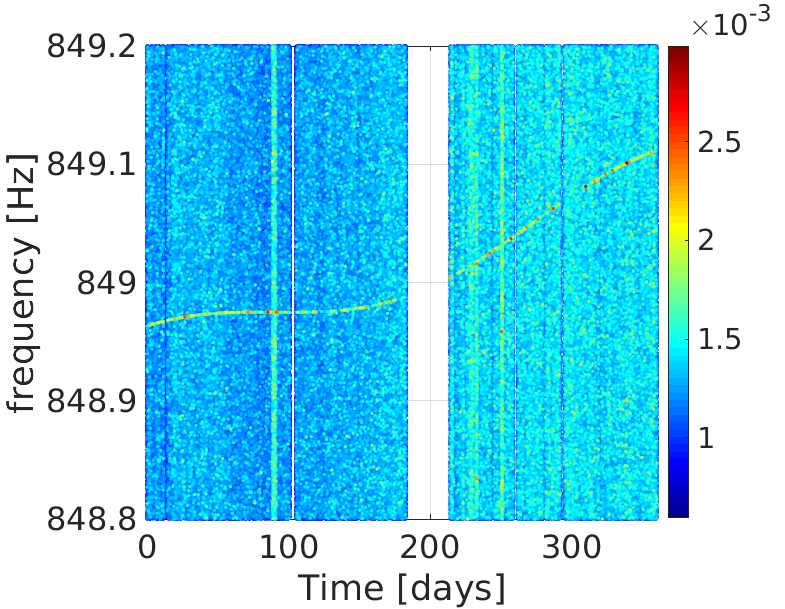}
\caption{Peakmap of H1 (left) and L1 (right) data showing one of the eight outliers removed by veto V5 in the second follow-up step, see \ref{tab:followup_1st_LL_LH}. All the 8 outliers were generated by the hardware injection ip1. A Doppler correction, with parameters not exactly equal with those of the signal, nevertheless aligns some of the signal peaks, thus producing an excess of counts in the \fh\ map.}
\label{fig:FH_peakmap_HL}
\end{figure*}

Table \ref{tab:followup_1st_LL_LH} summarize the results of the first follow-up stage over coincident H1 - L1 outliers, for each of the four analyzed frequency bands, given in the first column.
The second columns is the value of $T_{\rm FFT}$ used at this stage, $N_{\rm i}$ the initial number of outliers to which the follow-up is applied. Subsequent columns indicate the number of candidates removed by the various vetoes, indicated as $V_{\rm i}, i=1,..5$ and discussed in the section \ref{subse:fh_fu}. The last column shows the number of outliers surviving the first follow-up stage.
\begin{table}[h]
\begin{center}
\begin{tabular}{l r c c c c c c c c}
\hline
Band [Hz] & $T_{\rm FFT}$ [s] & $N_{\rm i}$ & V1 & V2 & V3 & V4 & V5 & $N_{\rm s}$\\
\hline \hline
$10-128$ & 24576 &  4007 & 3988  & 4& 0 & 2 &10&3 \\
$128-512$ & 24576&  12439 & 12422 &0 & 1& 13 & 3&0\\
$512-1024$ & 8192 & 10033 & 10017&1 & 0&5 &2 &8\\
$1024-2048$ & 8192 & 7440 & 7413 & 2 &0 & 2 &5&18\\
\hline
\end{tabular}
\caption[]{Main quantities regarding the first follow-up stage for H1 - L1 coincident outliers. $T_{\rm FFT}$ is the FFT duration used in the follow-up, $N_{\rm i}$ is the initial number of outliers to which the follow-up is applied, while $V_{\rm i}, i=1,..5$ indicate the number of outliers removed by the subsequent vetoes. $N_{\rm s}$ is the number of outliers surviving the first follow-up stage.}
\label{tab:followup_1st_LL_LH}
\end{center}
\end{table}
As it can be seen from the last column, 29 outliers survive this follow-up stage.
Tab. \ref{tab:fu_HLV} shows the same quantities for the follow-up of coincident H1 - Virgo and L1 - Virgo outliers, which have been selected in the lowest frequency band, from 10 to 128 Hz.
\begin{table}[ht]
\begin{center}
\begin{tabular}{l r c c c c c c c c c}
\hline
Detector&Band [Hz] & $T_{\rm FFT}$ [s]& $N_{\rm i}$& V1 & V2 & V3 & V4 & V5 &$N_{\rm s}$ \\
\hline
\hline
LL-AV & $10-128$ & 24576 & 1132 & 1127 & 4  & 0 & 0&1& 0 \\
LH-AV & $10-128$ & 24576 & 1143 & 1132 & 10 & 0 & 1&0&0\\
\hline
\end{tabular}
\caption[]{Main quantities regarding the first follow-up stage for H1 - Virgo (LH-AV) and L1 - Virgo (LL-AV)  coincident outliers. The shown quantities are the same as for Tab. \ref{tab:followup_1st_LL_LH}. The corresponding frequency band is 10-128 Hz.}
\label{tab:fu_HLV}
\end{center}
\end{table}
In this case, all the outliers have been discarded.
Outliers which survived the first follow-up stage have been analyzed with a second step based on the same procedure as before but with a further increase in the FFT duration, which has been roughly doubled. The main quantities for the second follow-up stage are shown in Tab. \ref{tab:followup_2nd_LL_LH}.
\begin{table}[h]
\begin{center}
\begin{tabular}{l r c c c c c c c c}
\hline
Band [Hz] & $Tfft[s]$& $N_{i}$ & V1 & V2 & V3 & V4 & V5 & $N_{s}$\\
\hline \hline
$10-128$ & 49152 & 3 & 3 & 0& 0 & 0 &0&0 \\
$512-1024$ & 16384 & 8& 0 &0 & 0&0 &8 &0\\
$1024-2048$ & 16384 & 18  & 16 & 0&0 &0&2&0\\
\hline
\end{tabular}
\caption[]{Main quantities regarding the second follow-up stage for H1 - L1 surviving outliers.}
\label{tab:followup_2nd_LL_LH}
\end{center}
\end{table}
The eight outliers in the band 512 - 1024 Hz are due to {\it hardware injection} ip1. An example is shown in Fig.  \ref{fig:FH_peakmap_HL}, where the peakmap after Doppler correction is plotted for a small frequency range around the outlier frequency.
Although the outlier parameters are relatively far from those of ip1, it is expected, especially in the case of a strong signal like this that - due to parameter correlations - outliers can spread over a rather large portion of the parameter space around the exact signal.

\subsubsection{Upper limits}
\begin{figure}
\centering
    \includegraphics[width=1.1\columnwidth]{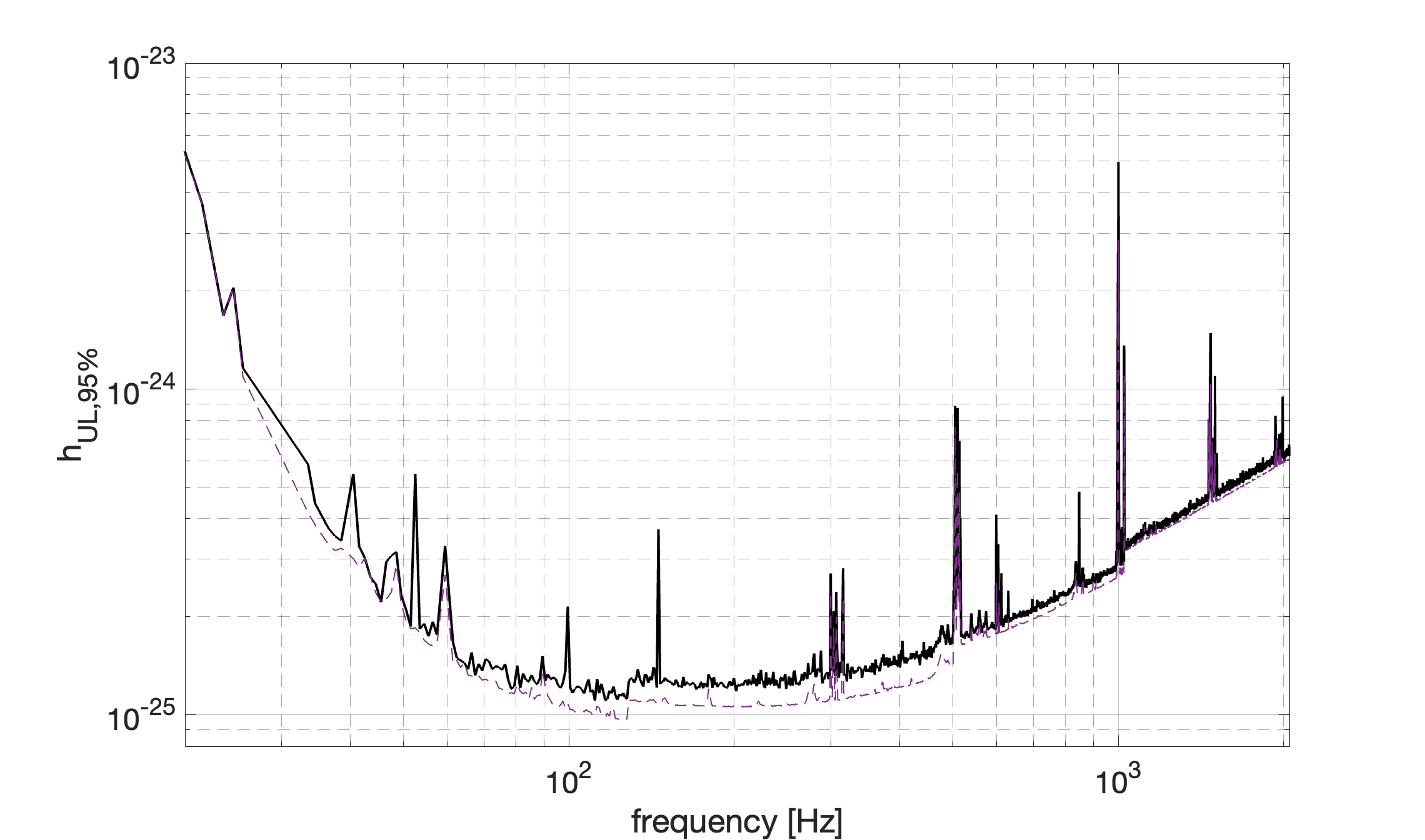}
    \caption{O3 conservative upper limit estimation (bold continuous curve) and sensitivity lower bound (light dashed curve) for the \fh\ search.}
    \label{fig:fhulresults}
\end{figure}

Having concluded that no candidate has a likely astrophysical origin, we have computed upper limits following the method described in Sec. \ref{sec:fhulmethod}. Results are shown in Fig. \ref{fig:fhulresults}. Although the search has been carried with a minimum frequency of 10 Hz, due to the unreliable calibration below 20 Hz, upper limits are given starting from this minimum frequency.
The bold continuous curve represents our conservative upper limit estimation, computed on 1 Hz sub-bands and based on the maximum CR, while the lighter dashed curve is a (non-conservative) lower bound, obtained using the minimum CR in each sub-band. We expect the search sensitivity, defined as the minimum detectable strain amplitude, to be comprised among the two curves. The minimum upper limit is about $1.1\times 10^{-25}$, at 116.5 Hz.

The search distance reach, expressed as a relation between the absolute value of the first frequency derivative and the frequency of detectable sources for various source distances, under the assumption the GW emission is the only spin-down mechanism (NSs in this case are often dubbed as {\it gravitars} \citep{2005MNRAS.359.1150P}), is shown in Fig.~\ref{fig:ell_vs_f}.

\subsubsection{Hardware Injections}

Table \ref{tab:HI_FH} shows the error of the recovered signal with respect to the hardware injections. The reported values have been obtained at the end of the first follow-up stage, which was enough to confidently detect the reported signals. The second column gives the total distance metric, defined in Sec. \ref{ssect:FHm}, among the injection and the corresponding strongest analysis candidate. Columns 3-6 give the error values for the individual parameters. Column 7 indicate the CR of the strongest candidate corresponding to each injection, and the last column gives the expected number of candidates due to noise, having the same (or bigger) CR value, after taking into account the trial factor.  
As shown in the Table, we have been able to detect 5 injections in the analyzed parameter space and the estimated
parameters do show a good agreement with the injected ones.  
All reported values are the mean of the values obtained separately for the Livingston and Hanford detectors, with the exception of the CR and $N_n$ for ip3, for which the reported values refer to Livingston alone. This hardware injection is in fact very weak and it was confidently detected, after the first follow-up stage, only in Livingston detector, which has a better sensitivity at the injection frequency. 
\begin{table*}[tbp]
\begin{center}
\begin{tabular}{l r c c c c c c c}
\hline
Injection & $d_\mathrm{FH}$ & $\Delta f$ [Hz] & $\Delta\dot{f}$ [nHz/s] & $\Delta\lambda$ [deg] & $\Delta\beta$ [deg] & CR & $N_{n}$\\
\hline \hline
ip1 & 0.77 & $1.15 \times 10^{-4}$ & $15.11 \times 10^{-3}$ & 0.015 & -0.027 & 51.76   &0 \\
ip3 & 1.05 & $4.88 \times 10^{-5} $ & $3.52 \times 10^{-3}$   & 0.088 & -0.377 & 6.34*   &0.04\\
ip5 & 1.92 & $2.65 \times 10^{-5}$ & $9.64 \times 10^{-3}$  & 0.615 & -0.130 &41.58    &0\\
ip6 & 0.16 & $9.27 \times 10^{-6}$ & $1.31 \times 10^{-3}$  & 0.009 &  0.045 &56.05    &0\\
ip14 & 1.52& $5.77 \times 10^{-4}$ & $52.27 \times 10^{-3}$  & 0.054 &  0.521 &20.58    &0\\
\hline
\end{tabular}
\caption[]{Hardware injection recovery by the {\fh} pipeline. The second column indicates the total distance metric among the injection and the corresponding strongest analysis candidate. Columns 3-6 give the error values for the individual parameters (frequency, spin-down, ecliptical longitude and latitude). Column 7 indicate the CR of the strongest candidate corresponding to each injection, and the last column gives the expected number of candidates due to noise, having the same (or bigger) CR value, after taking into account the trial factor. All the reported values are the mean of individual values found separately in Livingston and Hanford detectors, with the exception of the CR and $N_n$ for ip3, indicated by an asterisk, for which the reported values refer to Livingston alone. This hardware injection is very weak and it was confidently detected, after the first follow-up stage, only in Livingston, which has a better sensitivity at the injection frequency.}
\label{tab:HI_FH}
\end{center}
\end{table*}

\subsection{\sh}
\label{ssect:SHr}

\subsubsection{Candidate follow-up}

Table~\ref{tab:sh_outlier_per_stage} summarizes 
the number of outliers discarded by each of the veto and
follow-up stages employed in this search. 
A total of 36 candidates survive the complete suite of veto and follow-up stages of the \sh{} pipeline.
Candidates can be grouped into two sets according to their corresponding $\mathcal{F}$-statistic value:
31 candidates present a value of $2\tilde{\mathcal{F}} \sim \mathcal{O}(10^3)$,
while the remaining 5 candidate only achieve $2\tilde{\mathcal{F}} \sim \mathcal{O}(30)$.
Their corresponding parameters are collected in Table~\ref{tab:sh_surviving_candidates}.

The 31 strong candidates present consistent values with the only two hardware injections within the \sh{} search range:
24 candidates are ascribed to the hardware injection ip0,
while 7 candidates are ascribed to the hardware injection ip3.
Parameter deviation of the loudest candidate associated to each injection are reported in Table~\ref{tab:sh_hi}.

\begin{table}[]
    \centering
    \begin{tabular}{lrr}
        \toprule
            Search stage  &  Candidates & \% removed \\
        \midrule
            Clustering & 456000 &\\
            Line veto & 414459 & 9\%\\
            2$\mathcal{F}$ threshold & 3767 & 99\% \\
            Stage 0 v.s. Stage 1 & 697 & 18\%\\
            Stage 1 v.s. Stage 2 & 172 & 75\% \\
            Stage 3 v.s. Stage 3 & 90 & 48\%\\
            Stage 3 v.s. Stage 4 & 48 & 47\%\\
            Stage 4 v.s. Stage 5 & 36  &25\%\\
        \bottomrule
    \end{tabular}
    \caption{Summary of candidates processed by each of the veto and follow-up stages of the \sh{} search.}
    \label{tab:sh_outlier_per_stage}
\end{table}

\begin{table*}
\begin{center}
    \begin{tabular}{cccccccc}
\toprule
 Band &  Candidate &         $f$ [Hz] &  $\dot{f}$ [nHz/s] &    $\alpha$ [rad] &     $\delta$ [rad] & $2\tilde{\mathcal{F}}$ & Comment \\
\midrule
  834 &          4 &  85.872761414 &  2.41584$\cdot 10^{-3}$ & 3.143782737 &  1.165116066 &   30.54 &  Broad spectral feature in H1 \\
  834 &          9 &  85.873653124 & -9.35774$\cdot 10^{-2}$ & 3.409549407 &  1.385107830 &   36.25 &  Broad spectral feature in H1 \\
 1227 &         35 &  95.697667346 & -4.89489$\cdot 10^{-2}$ & 1.593327050 & -1.292111453 &   31.53 & Narrow spectral feature in H1 \\
 1229 &          5 &  95.725474979 & -9.63949$\cdot 10^{-1}$ & 0.260240661 & -1.008336167 &   30.87 & Narrow spectral feature in H1 \\
 1754 &          1 & 108.857159405 & -8.04825$\cdot 10^{-7}$ & 3.113189707 & -0.583577133 & 1055.70 &        Hardware injection ip3 \\
 1754 &          2 & 108.857159406 & -8.29209$\cdot 10^{-7}$ & 3.113189734 & -0.583577139 & 1055.69 &        Hardware injection ip3 \\
 1754 &          5 & 108.857159404 & -7.43862$\cdot 10^{-7}$ & 3.113189647 & -0.583577277 & 1055.71 &        Hardware injection ip3 \\
 1754 &         10 & 108.857159405 & -7.92726$\cdot 10^{-7}$ & 3.113189663 & -0.583577189 & 1055.71 &        Hardware injection ip3 \\
 1754 &         13 & 108.857159406 & -8.38377$\cdot 10^{-7}$ & 3.113189745 & -0.583577097 & 1055.69 &        Hardware injection ip3 \\
 1754 &         14 & 108.857159405 & -8.14434$\cdot 10^{-7}$ & 3.113189656 & -0.583577155 & 1055.69 &        Hardware injection ip3 \\
 1754 &         34 & 108.857159404 & -7.09929$\cdot 10^{-7}$ & 3.113189613 & -0.583577327 & 1055.69 &        Hardware injection ip3 \\
 7251 &         10 & 246.297680589 & -2.24806$\cdot 10^{-2}$ & 1.425124776 & -1.242786654 &   35.79 & Narrow spectral feature in H1 \\
 8022 &          0 & 265.575086278 & -4.14962$\cdot 10^{-3}$ & 1.248816426 & -0.981180252 & 1543.70 &        Hardware injection ip0 \\
 8022 &          1 & 265.575086279 & -4.14969$\cdot 10^{-3}$ & 1.248816468 & -0.981180265 & 1543.68 &        Hardware injection ip0 \\
 8022 &          2 & 265.575086278 & -4.14961$\cdot 10^{-3}$ & 1.248816419 & -0.981180239 & 1543.69 &        Hardware injection ip0 \\
 8022 &          3 & 265.575086278 & -4.14964$\cdot 10^{-3}$ & 1.248816434 & -0.981180252 & 1543.69 &        Hardware injection ip0 \\
 8022 &          4 & 265.575086278 & -4.14964$\cdot 10^{-3}$ & 1.248816444 & -0.981180252 & 1543.70 &        Hardware injection ip0 \\
 8022 &          5 & 265.575086277 & -4.14958$\cdot 10^{-3}$ & 1.248816405 & -0.981180243 & 1543.70 &        Hardware injection ip0 \\
 8022 &          7 & 265.575086279 & -4.14968$\cdot 10^{-3}$ & 1.248816456 & -0.981180263 & 1543.69 &        Hardware injection ip0 \\
 8022 &         28 & 265.575086278 & -4.14965$\cdot 10^{-3}$ & 1.248816441 & -0.981180257 & 1543.69 &        Hardware injection ip0 \\
 8023 &          0 & 265.575086278 & -4.14964$\cdot 10^{-3}$ & 1.248816439 & -0.981180255 & 1543.70 &        Hardware injection ip0 \\
 8023 &          1 & 265.575086278 & -4.14961$\cdot 10^{-3}$ & 1.248816417 & -0.981180250 & 1543.70 &        Hardware injection ip0 \\
 8023 &          3 & 265.575086278 & -4.14966$\cdot 10^{-3}$ & 1.248816464 & -0.981180249 & 1543.68 &        Hardware injection ip0 \\
 8023 &          4 & 265.575086279 & -4.14969$\cdot 10^{-3}$ & 1.248816466 & -0.981180264 & 1543.68 &        Hardware injection ip0 \\
 8023 &          7 & 265.575086279 & -4.14967$\cdot 10^{-3}$ & 1.248816448 & -0.981180256 & 1543.69 &        Hardware injection ip0 \\
 8023 &          8 & 265.575086279 & -4.14966$\cdot 10^{-3}$ & 1.248816453 & -0.981180260 & 1543.71 &        Hardware injection ip0 \\
 8023 &          9 & 265.575086278 & -4.14963$\cdot 10^{-3}$ & 1.248816431 & -0.981180254 & 1543.70 &        Hardware injection ip0 \\
 8023 &         10 & 265.575086275 & -4.14945$\cdot 10^{-3}$ & 1.248816284 & -0.981180203 & 1543.26 &        Hardware injection ip0 \\
 8023 &         11 & 265.575086278 & -4.14962$\cdot 10^{-3}$ & 1.248816419 & -0.981180255 & 1543.69 &        Hardware injection ip0 \\
 8023 &         12 & 265.575086278 & -4.14963$\cdot 10^{-3}$ & 1.248816435 & -0.981180249 & 1543.70 &        Hardware injection ip0 \\
 8023 &         13 & 265.575086277 & -4.14956$\cdot 10^{-3}$ & 1.248816392 & -0.981180234 & 1543.66 &        Hardware injection ip0 \\
 8023 &         14 & 265.575086278 & -4.14966$\cdot 10^{-3}$ & 1.248816450 & -0.981180252 & 1543.70 &        Hardware injection ip0 \\
 8023 &         16 & 265.575086278 & -4.14962$\cdot 10^{-3}$ & 1.248816403 & -0.981180252 & 1543.65 &        Hardware injection ip0 \\
 8023 &         18 & 265.575086278 & -4.14962$\cdot 10^{-3}$ & 1.248816430 & -0.981180248 & 1543.66 &        Hardware injection ip0 \\
 8023 &         19 & 265.575086278 & -4.14963$\cdot 10^{-3}$ & 1.248816436 & -0.981180254 & 1543.72 &        Hardware injection ip0 \\
 8023 &         34 & 265.575086278 & -4.14965$\cdot 10^{-3}$ & 1.248816452 & -0.981180250 & 1543.72 &        Hardware injection ip0 \\
\bottomrule
\end{tabular}

\end{center}
\caption{Surviving candidates of the \sh{} multi-stage MCMC follow-up using \texttt{PyFstat}. 
$2\tilde{\mathcal{F}}$ corresponds to the loudest fully-coherent $\mathcal{F}$-statistic value of the MCMC run. 
Band index corresponds to a frequency of $(65 + 0.025 \times \textrm{Band})$ Hz. Reference time is GPS 1238166018.}
\label{tab:sh_surviving_candidates}
\end{table*}

\begin{table*}
\begin{center}
\begin{tabular}{ccrrrrrr}
\toprule
Injection & $2\tilde{\mathcal{F}}$ & $\Delta f$ [Hz] & $\Delta \dot{f}$ [nHz/s] & $\Delta \alpha$ [rad] & $\Delta \delta$ [rad] & $\Delta \alpha$ [deg] & $\Delta \delta$ [deg] \\
\midrule
ip0 &  1543.72 & $-4.80 \times 10^{-9}$ & $3.52 \times 10^{-7}$ & $-2.82 \times 10^{-7}$ & $-2.49 \times 10^{-8}$ & $-1.62\times 10^{-5}$ & $-1.43 \times 10^{-6}$\\
ip3 & 1055.71  & $1.16 \times 10^{-8}$ & $-7.29 \times 10^{-7}$ & $9.35 \times 10^{-7}$  & $1.53 \times 10^{-6}$ & $5.35\times 10^{-5}$ & $8.74\times 10^{-5}$\\
\bottomrule
\end{tabular}
\caption{Hardware injection recovery by the \sh\ pipeline. 
For each hardware injection within search range we report the dimension-wise errors with respect to loudest surviving candidate of the follow-up.}
\label{tab:sh_hi}
\end{center}
\end{table*}

The five weaker candidates are manually inspected using 
the segment-wise $\mathcal{F}$-statistic on 660 coherent segments,
in a similar manner to that in~\cite{PhysRevD.104.082004, Tenorio:2021njf}.

\begin{figure}
    \centering
    \includegraphics[width=\columnwidth]{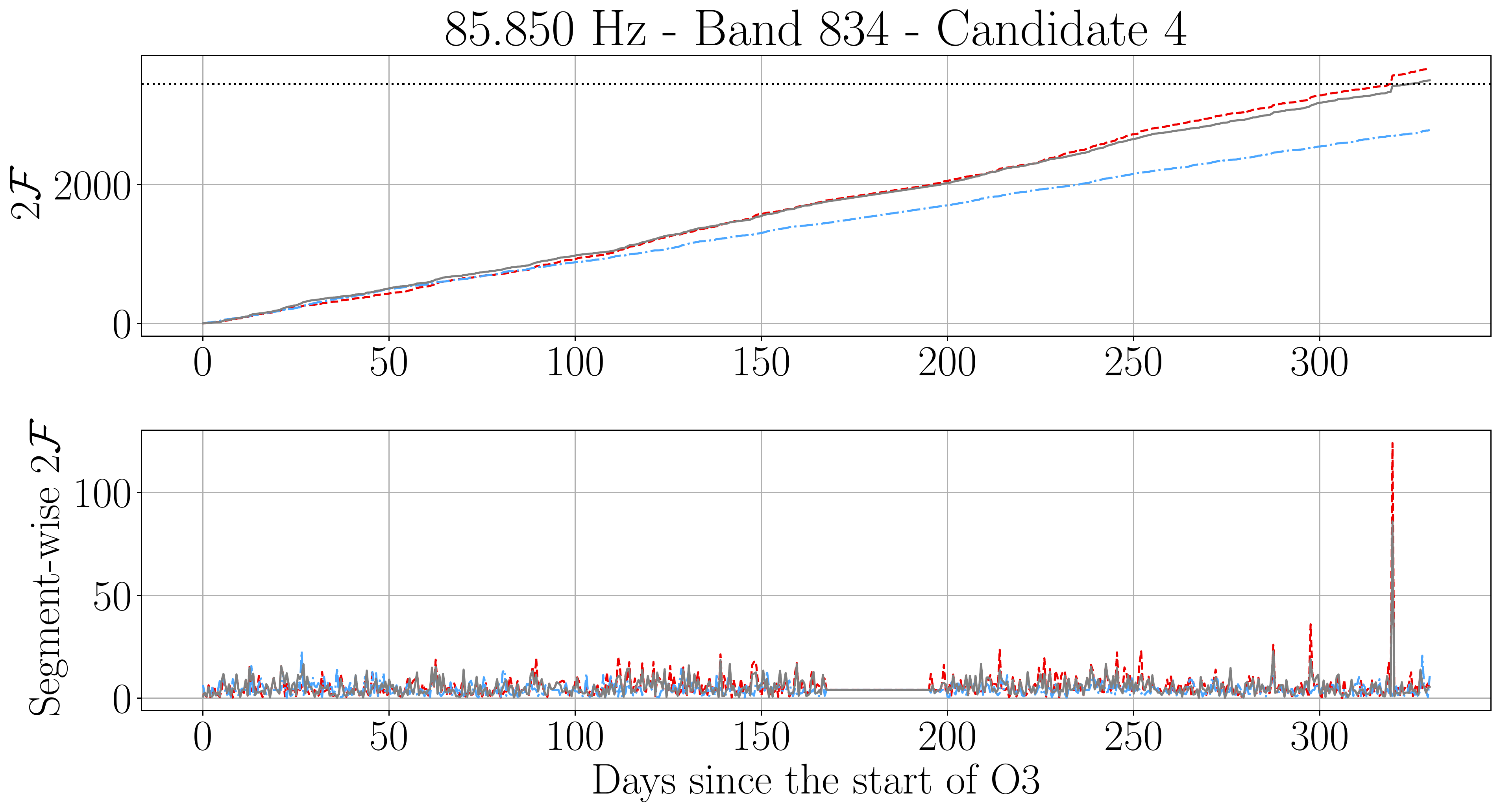}
    \includegraphics[width=\columnwidth]{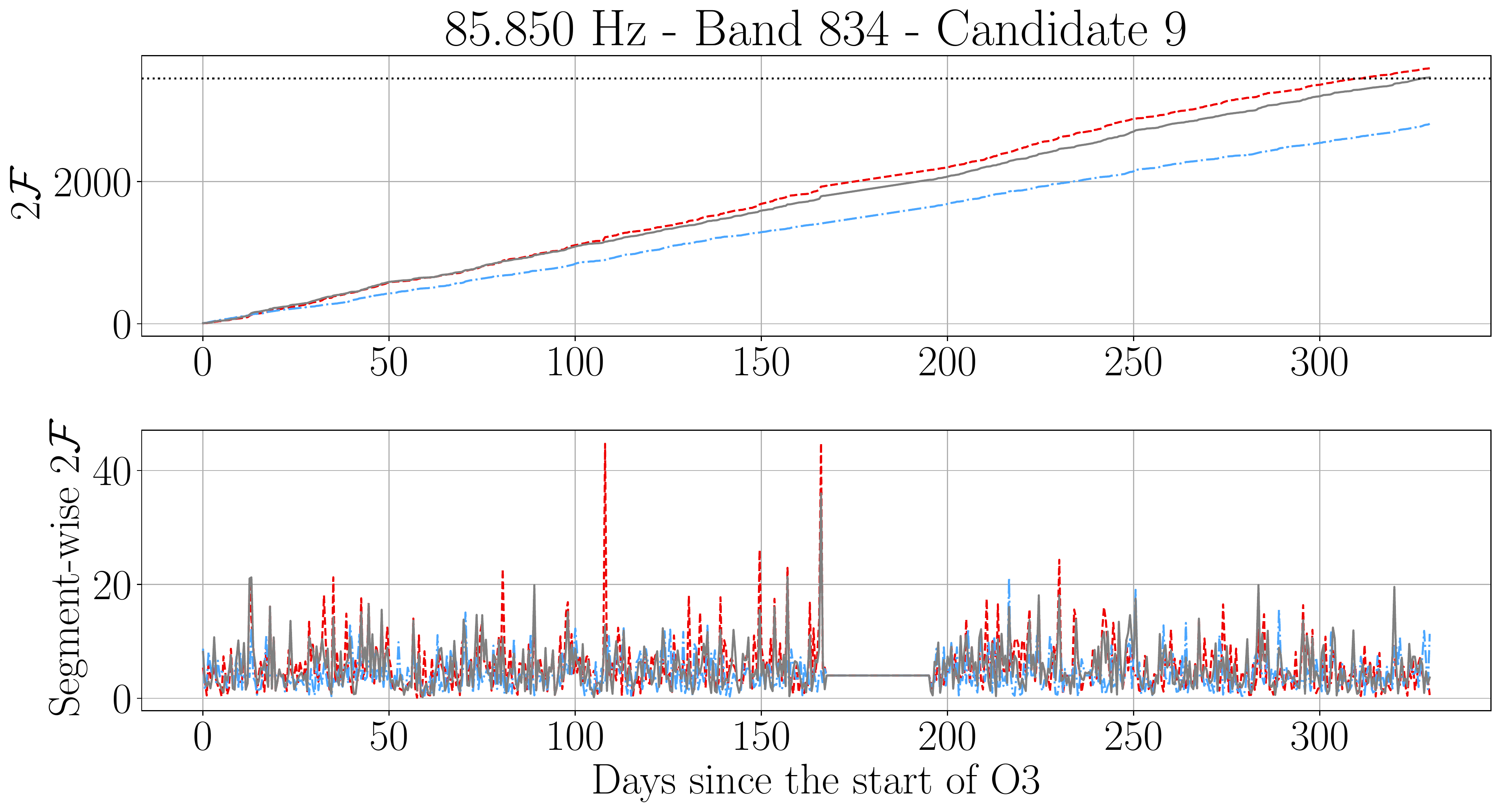}
    \caption{\sh{} candidates consistent with a broad spectral artifact 
    in the H1 detector. Upper panel shows the cumulative semicoherent
    $\mathcal{F}$-statistic using 660 coherent segments ($\Tcoh=0.5$ days).
    Lower panel shows the segment-wise $\mathcal{F}$-statistic. 
    Dashed red line represents the single-detector $\mathcal{F}$-statistic 
    using H1-only data; 
    dot-dashed blue line represents the  single-detector
    $\mathcal{F}$-statistic using L1-only data. Solid gray line represents
    the multi-detector $\mathcal{F}$-statistic. Dotted horizontal line
    represents the threshold of $2\mathcal{F}=3450$ set at the initial
    follow-up stage.}
    \label{fig:sh_outliers_85}
\end{figure}

The first pair of candidates is found around \mbox{85.850 Hz}, where the H1 detector presents a broad spectral feature.
As shown in Fig~\ref{fig:sh_outliers_85}, their single-detector \mbox{$\mathcal{F}$}-statistic is more prominent in the H1 detector rather than the L1 detector, 
and scores over the multi-detector \mbox{$\mathcal{F}$}-statistic. 
These characteristics point towards an instrumental, rather than astrophysical, origin.

\begin{figure}
    \centering
    \includegraphics[width=\columnwidth]{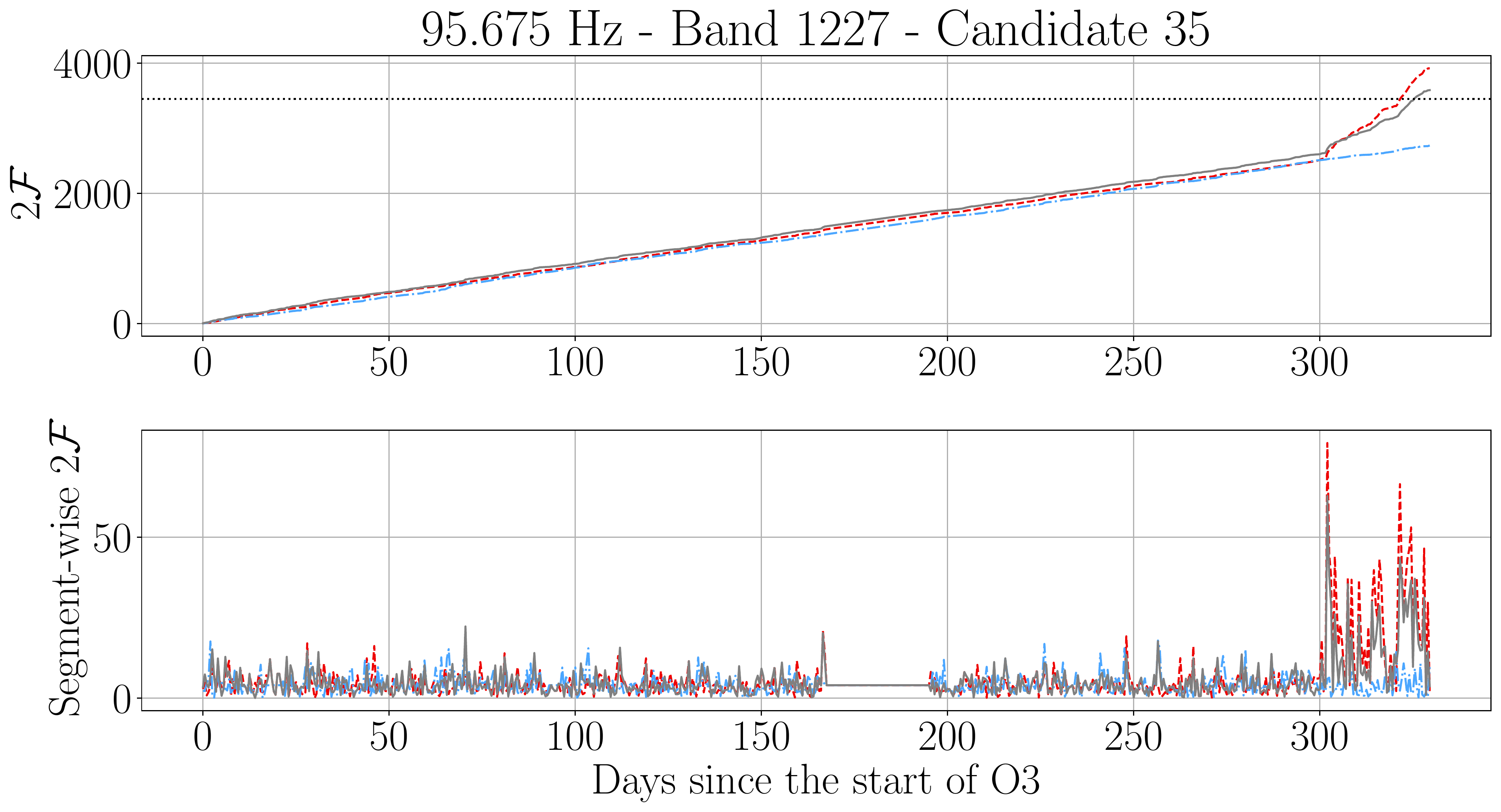}
    \includegraphics[width=\columnwidth]{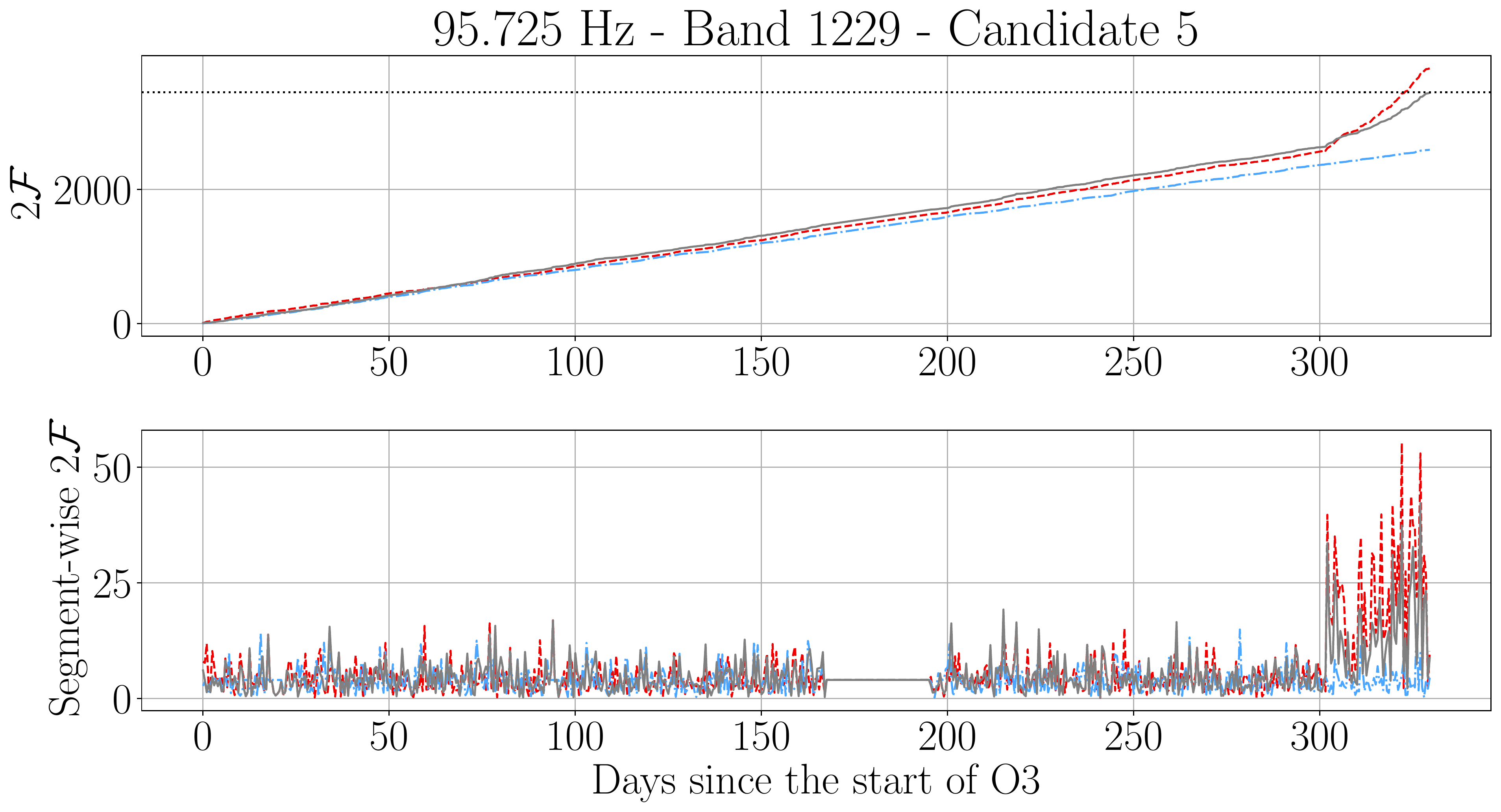}
    \caption{\sh{} candidate consistent with two narrow spectral artifacts of unknown origin in the H1 detector. The legend is equivalent to that of
    Fig.~\ref{fig:sh_outliers_85}.}
    \label{fig:sh_outliers_95}
\end{figure}

A second pair of candidates is found around \mbox{95.7 Hz}. This frequency band is populated by narrow spectral artifacts of unknown origin in the H1 detector. 
Correspondingly, as shown in Fig.~\ref{fig:sh_outliers_95}, the single-detector \mbox{$\mathcal{F}$} statistic is prominent in the H1 detector rather than the L1 detector.
Due to the narrowness of the feature, 
in this case the accumulation is better localized around a fraction of the run.
As in the previous case, the single-detector $\mathcal{F}$-statistic scores over the multi-detector \mbox{$\mathcal{F}$}-statistic. 
These characteristics point towards an instrumental origin.

The last weak candidate in the vicinity of 246.275 Hz,
where the H1 detector presents another narrow spectral
artifact of unknown origin. 
The single-detector $\mathcal{F}$-statistic is
more prominent in the H1 detector than in the L1 detector,
and accumulates rapidly at the beginning of the run.
As in the previous cases, this behavior is consistent with
that of an instrumental artifact.

This concludes the analysis of surviving candidates of
the \sh{} pipeline. Every single one of them could be 
related to an instrumental feature.

\begin{figure}
    \centering
    \includegraphics[width=\columnwidth]{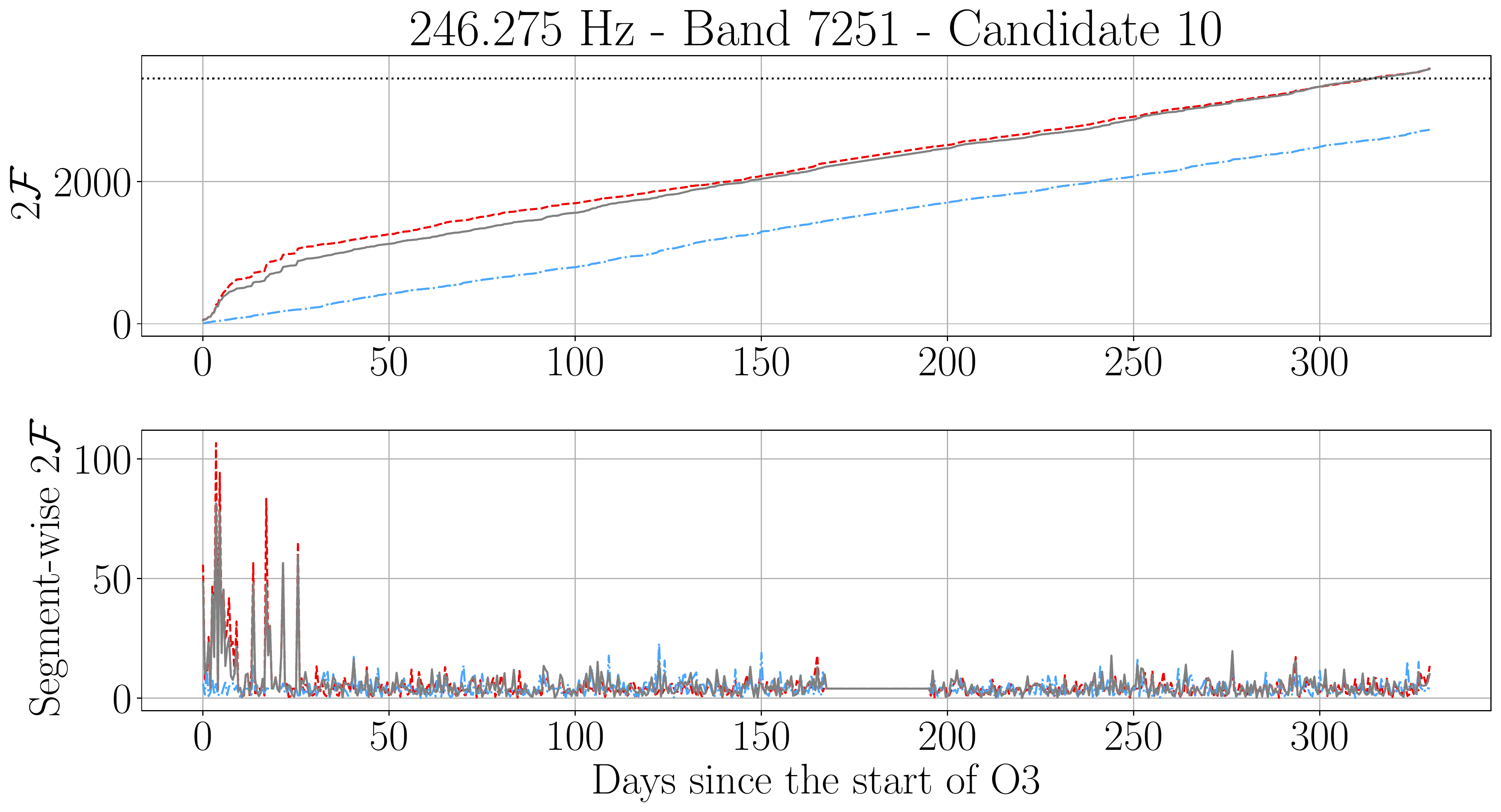}
    \caption{\sh{} candidate consistent with a narrow spectral artifact of unknown origin in the H1 detector. The legend is equivalent to that of
    Fig.~\ref{fig:sh_outliers_85}.}
    \label{fig:sh_outliers_246}
\end{figure}

\subsubsection{Sensitivity estimation}

\begin{figure}
    \centering
    \includegraphics[width=\columnwidth]{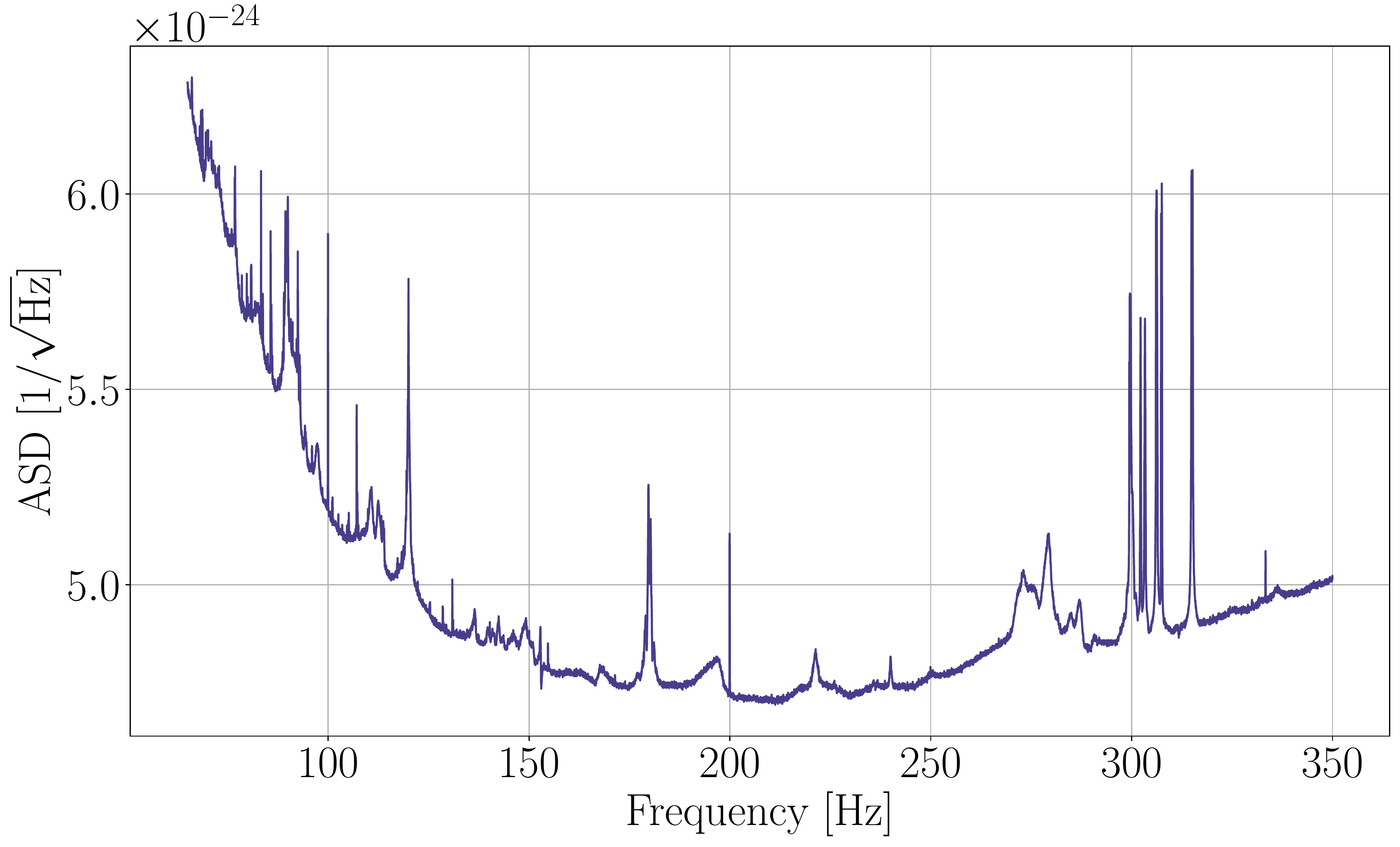}
    \caption{ASD employed by the \sh{} pipeline to estimate the sensitivity of the search.
    ASD is computed as the square root of the single-sided inverse-square averaged PSD using data from both
    the H1 and L1 advanced LIGO detectors, as explained in the text surrounding Eq.~\eqref{eq:sh_PSD}.
    }
    \label{fig:sh_asd}
\end{figure}

We estimate the search sensitivity following the same procedure as previous searches
\cite{2017PhRvD..96f2002A, 2018PhRvD..97j2003A, 2019PhRvD.100b4004A, Covas:2020nwy, LIGOScientific:2020qhb}.
Search sensitivity is quantified using the \emph{sensitivity depth}~\cite{2015PhRvD..91f4007B, Dreissigacker:2018afk}
\begin{equation}
    \D{} = \frac{\sqrt{S_{\textrm{n}}}}{h_0}\;,
    \label{eq:depth}
\end{equation}
where $S_{\textrm{n}}$ represents the power spectral density (PSD) of the data, computed
as the inverse squared average of the individual SFT's running-median PSD~\cite{2008PhRvD..77b2001A, LIGOScientific:2020qhb}
\begin{equation}
    S_{\textrm{n}}(f) = \sqrt{\frac{N_{\alpha}}{\sum_{\alpha} \left[S_{\alpha}(f)\right]^{-2}}} \;.
    \label{eq:sh_PSD}
\end{equation}
where $S_{\alpha}$ represents the running-median noise floor estimation using 101 bins from the SFT labeled by starting time
$t_{\alpha}$ (including SFTs from both the H1 and L1 detectors)  and $N_{\alpha}$ represents the total number of SFTs. 
The resulting amplitude spectral density (ASD) $\sqrt{S_{\textrm{n}}}$ is shown in Fig.~\ref{fig:sh_asd}.

\begin{figure}
    \centering
    \includegraphics[width=\columnwidth]{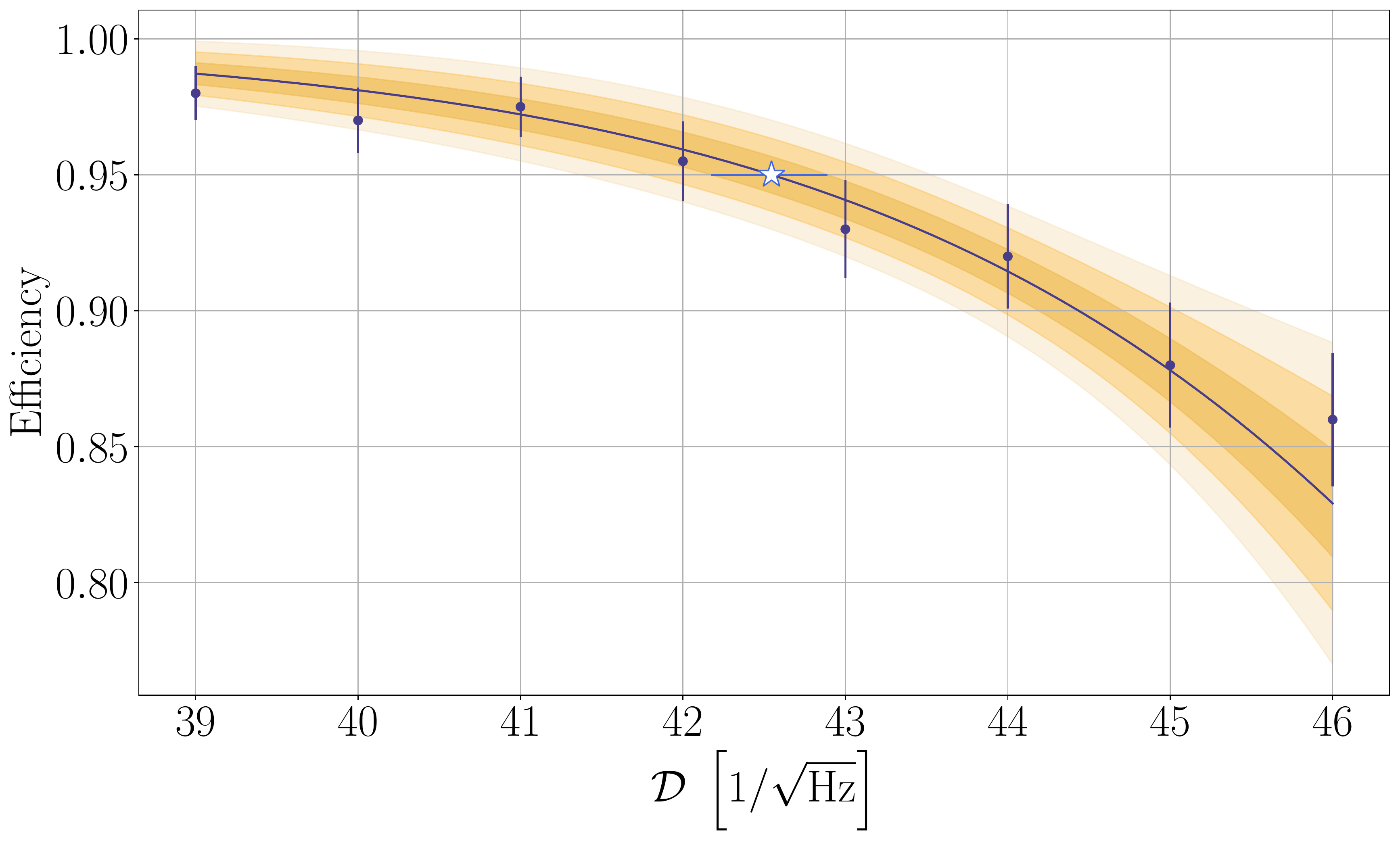}
    \caption{Example computation of $\Dul$ (white star) at a frequency band  by fitting a 
    sigmoid function (blue solid line) to a set of efficiencies (blue dots) computed using 200 injections 
    at each sensitivity depth for the {\sh} search. Shaded regions represent 1, 2, and 3 sigma envelopes of
    the sigmoid fit. Error bars are computed as discussed in the main text.}
    \label{fig:sh_efficiency_example}
\end{figure}

The sensitivity depth $\Dul{}$ corresponding to a 95\% average detection rate is characterized 
by adding a campaign of software-simulated signals into the data. 
Simulated signals are added into 150 representative frequency bands at several sensitivity depth values
bracketing the $\Dul{}$ value in each band, as represented in Fig.~\ref{fig:sh_efficiency_example}.
For each sensitivity depth, 200 simulated signals drawn from uniform distribution in phase and amplitude 
parameters are added into the data. The \sh{} is run on each of these signals in order to evaluate how
many of them are detected, and the resulting toplists are clustered using the same configuration as in the
main stage of the search.

For each simulated signal, we retrieve the best forty resulting clusters. 
The following two criteria must be fulfilled in order to label a simulated signal as ``detected''. 
First, the loudest significance of at least one of the selected clusters must be higher than the minimum
significance recovered by the corresponding all-sky clustering;
this ensures the signal is significant enough to be selected for a follow-up stage. 
Second, the parameters of the loudest candidate in said clusters must be closer than two parameter-space
bins (see Eq.~\eqref{eq:sh_resolution} and Table~\ref{tab:sh_resolution}) from the simulated-signal's parameter,
as otherwise the follow-up would have missed the signal.

The efficiency associated to each sensitivity depth $E$ is computed as the fraction of simulated signals 
labeled as detected. A binomial uncertainty $\delta E$ is associated to each efficiency
\begin{equation}
    \label{binomialerror}
    \delta E = \sqrt{\frac{E \cdot (1-E)}{N_{\textrm{I}}}} \;,
\end{equation}
where $N_{\textrm{I}} = 200$ represents the number of signals. Then, we use \texttt{scipy}'s 
\texttt{curve\_fit} function~\cite{2020SciPy-NMeth} to fit a sigmoid curve to the data given by
\begin{equation}
    S(\D; a, b) = 1 - \frac{1}{1 + \exp{(-a \D + b)}}
\end{equation}
where $a, b$ represent the parameters to adjust. After fitting, this expression can be numerically
inverted to obtain $\Dul{}$. The uncertainty associated to the fit is compute through the covariance
matrix $C$ as
\begin{equation}
    \delta \Dul{} = \left. 
    \sqrt{\nabla_{a, b} S^{\textrm{T}} \cdot C \cdot  \nabla_{a, b} S}
    \right|_{\D = \Dul{}}
\end{equation}
where $\nabla_{a, b}$ represents the gradient with respect to the fitting parameters. This procedure
is exemplified in Fig.~\ref{fig:sh_efficiency_example}.
 
\begin{figure}
    \centering
    \includegraphics[width=\columnwidth]{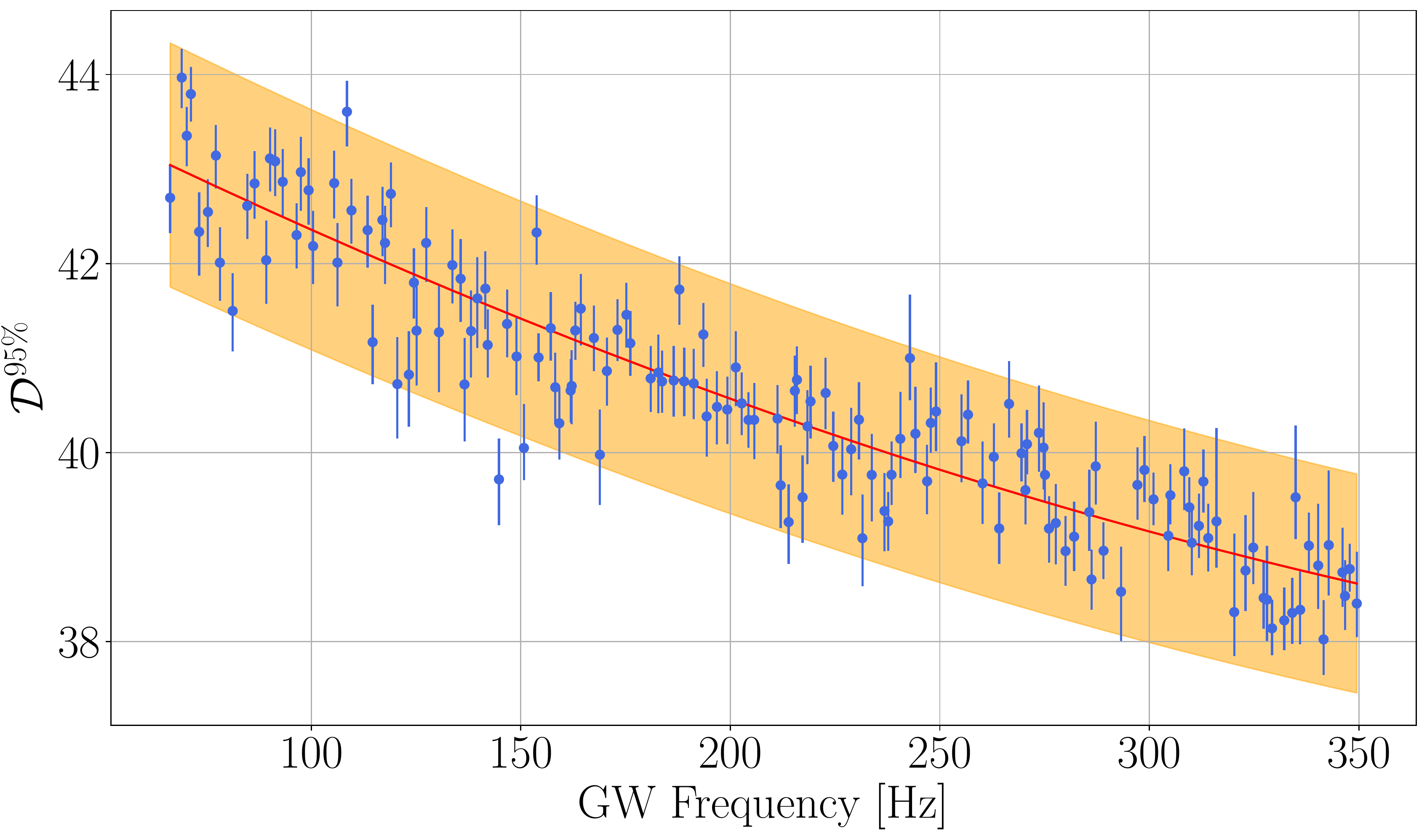}
    \caption{Wide-band interpolation $\Dul(f)$ of the results obtained by the \sh{} pipeline. 
    Each dot represents a $\Dul{}$ at a particular frequency band computed using 
    the procedure exemplified in Fig.~\ref{fig:sh_efficiency_example}. The red solid line represents
    a non-parametric interpolation using a Gaussian process regression, as discussed in the main text.
    The shaded region represents a 3\% relative error with respect to the interpolation and corresponds 
    to the 98\% credible interval.}
    \label{fig:sh_depth_fit}
\end{figure}

\begin{figure}
    \centering
    \includegraphics[width=\columnwidth]{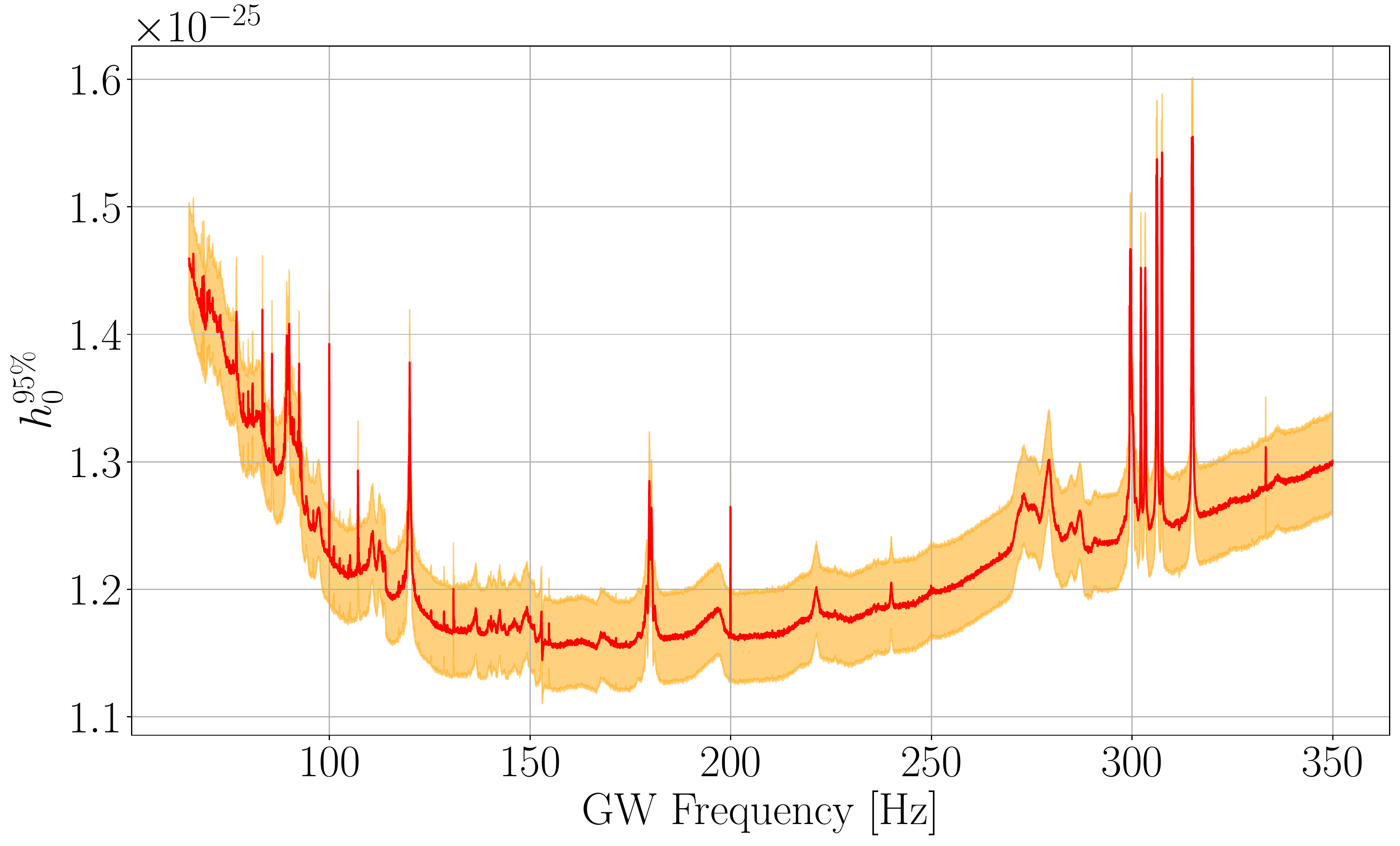}
    \caption{CW amplitude $\hul$ corresponding to the 95\% detection efficiency depth along
    the frequency band analyzed by the \sh{} pipeline. Solid line represents the implied $\hul$ from
    the wide-band $\Dul{}$ interpolation shown in Fig.~\ref{fig:sh_depth_fit}. Shaded region corresponds
    to the 3\% relative error with respect to the interpolation.}
    \label{fig:sh_h0_95}
\end{figure}

We compute the average wide-band $\Dul(f)$ value using Gaussian process regression, 
as shown in Fig.~\ref{fig:sh_depth_fit}. 
We fit a Gaussian process using to the ensemble of $\Dul{}$ obtained from the injection campaign 
using \texttt{scikit-learn}'s \texttt{GaussianProcessRegressor} with an \texttt{RBF} kernel~\cite{scikit-learn}. 
The uncertainty associated to the fit is computed as the 98\% credible region of the deviations with
respect to the Gaussian process regression, which corresponds to a 3\% relative uncertainty.
Equation~\eqref{eq:depth} allows us to translate $\Dul(f)$ into a corresonding CW amplitude $\hul(f)$,
shown in Fig.~\ref{fig:sh_h0_95}.

\subsection{\tdfstat}
\label{ssect:FStatr}

In the frequency bandwidth of [20, 750]~Hz that we analyze we have 3245 sub-bands that are 0.25~Hz wide and that are overlapped by 0.025~Hz.  
104 sub-bands were not analyzed because of the excessive noise originating mainly from the 1st harmonic of the violin mode,
1st and 2nd harmonics of the beam splitter violin mode, and 60~Hz mains line and its harmonics.
This leads to the loss of around 23.50~Hz of the band. Moreover, we have vetoed lines identified by the detector characterization group.
This leads to an additional 34.18~Hz band loss. Thus altogether 57.68~Hz of the band was vetoed, which constitutes 7.9\% of the 730~Hz band analyzed.
Consequently we searched 3141 sub-bands. For each sub-band we analyzed coherently 41 six-day time segments with the $\F$-statistic.
As a result with our $\F$-statistic threshold of 15.5 we obtained $5.47\times10^{10}$ candidates.

In the second stage of the analysis for each sub-band we search for coincidences among the candidates from the 41 time-domain segments.
For each sub-band and each hemisphere we find the candidate with the smallest coincidence false alarm probability, i.e.\ the most significant candidate.
As a result we have 6282 top candidates from our search. Among the top candidates we consider a candidate to be statistically significant
if the coincidence false alarm probability is less than 1\%. This leads to the selection of 311 candidates that we call {\em outliers}.
The outliers were subject to further investigation to determine whether they can be considered as true GW events.
Three of the outliers were determined to be `artificial' GW signals injected in hardware to the LIGO detectors data.

\subsubsection{Hardware injections}
\label{ssec:fstat_HI}

In the parameter space analysed by Time-Domain $\F$-statistic only six hardware injections were present.
These are injections ip0, ip2, ip3, ip5, ip10, and ip11.
In Table \ref{tab:tdfstat_hi} we have compared the parameters of the top candidates obtained in our search in the frequency sub-bands,
where the injections were made, with the parameters of the injections.
In the table we show the false alarm probability of coincidence of the top candidates
and the difference between the parameters of the candidate and the parameters of the injections.
\begin{table*}[tbp]
\begin{center}
\begin{tabular}{l r@{\hspace{3ex}} c@{\hspace{3ex}} c@{\hspace{3ex}} r r}
\hline
Injection & \ FAP \ & $\Delta f$ [Hz] & $\Delta\dot{f}$ [nHz/s] & $\Delta\delta$ [deg] & $\Delta\alpha$ [deg] \\
\hline \hline
ip10 & $ < 10^{-8} $ & $7.73 \times 10^{-4}$ & $3.91 \times 10^{-3}$   & 0.74  & 1.51 \\
ip11 & $ < 10^{-8} $ & $6.20 \times 10^{-3}$ & $7.22 \times 10^{-2}$ &  14.78 & 248.13 \\
ip5 & $ < 10^{-8} $ & $1.62 \times 10^{-4}$ & $6.64 \times 10^{-2}$  &  30.24 & 24.83 \\
ip3 & $ 0.997 $ & $5.21 \times 10^{-2}$ & $5.21 \times 10^{-2}$  & 3.85 & 1.56 \\
ip0 & $ 0.055 $ & $1.84 \times 10^{-1}$   & $1.15 \times 10^{-1}$  & 18.94 & 20.98 \\
ip2 & $ 0.275 $  & $4.41 \times 10^{-5}$    &$9.28 \times 10^{-3}$  & 1.12 & 0.074 \\
\hline
\end{tabular}
\caption[]{Hardware injection recovery with the Time-Domain $\F$-statistic pipeline.
The first column is the injection index ipN, where N is the injection number.
The last 4 columns are the differences between the true values of the parameters of the injected signal ipN
and the parameters of the most significant candidate in the sub-band where injection is added.
The second column is the false alarm probability associated with the topmost candidate.}
\label{tab:tdfstat_hi}
\end{center}
\end{table*}
We see that the two injections ip5 and ip10 are detected with a very high confidence.
Their false alarm probability is close to  0 and the errors in the parameter estimation are small.
The top candidate in the band where injection ip11 is located has a very small false alarm probability;
however, the right ascension of the candidate differs very much from the true value the right ascension of the injection.
A close analysis shows that this candidate is associated with a strong line present in the Hanford detector.
The line frequency is different from the hardware injection frequency by only around 10~mHz.
The amplitude of the injection ip11 is very low.
Its SNR in the 6-day segments that we analyse coherently with $\F$-statistic is around 4.
This is considerably lower than our threshold SNR of around 5.2 and it is not surprising that the injection is not recovered.
For the remaining 3 bands we see that the top candidates have parameters very close to the parameters of the hardware injections ip0, ip2 and ip3,
however their false alarm probabilities are greater than 1\% and we cannot consider these injections as detected.
The SNRs of the two detected injections ip5 and ip10 is considerably above our threshold of 5.2,
whereas SNRs of the 3 remaining injections ip0, ip2, and ip3 are close to our threshold and they could not be detected.

\subsubsection{Outliers}
\label{ssec:fstat_outliers}

We have identified {\bf 311} outliers in our search.
For these outliers the probability of being due to accidental coincidence between the candidates from the 41 time segments is less than 1\%. 

In our search we have vetoed the lines of known origin identified in LIGO detectors.
However, the LIGO data contained additional lines and interferences.
In order to identify the origin of the outliers in our search we have performed three independent investigations.
Firstly we compared our outliers with the lines of unknown origin identified by the LIGO data characterization group.
Secondly we have performed an independent search for strictly periodic signals in all the 6-day time-domain segments that we analyzed in our search.
We have searched for periodic signals separately in the data from the Hanford and the Livingston LIGO detector.
Thirdly we have performed a visual inspection of the outliers by searching the data with $\F$-statistic around the outliers separately in the two LIGO detectors.
In addition we have checked whether outliers are around the frequencies associated with the suspension violin mode 1st harmonic around 500~Hz
and the beam splitter violin mode 1st and 2nd harmonics around 300~Hz and 600~Hz respectively. 
As a result of the above study {\bf 204} outliers were found to be associated with lines and interferences present in the detector.
They were classified as follows. 146 originated from the Hanford detector, 21 were associated with the Livingston detector.
One line that appeared in both detectors was the 20~Hz tooth of the 1~Hz comb known to be present in both detectors.
36 outliers were associated with the two violin mode resonances.

{\bf 2} outliers are pulsar injections ip5 and ip10 that were confidently detected and they are described in Sec.\ \ref{ssec:fstat_HI}.

{\bf One} of the outliers was associated with the pulsar injection ip6.
The frequency of the outlier was only 15~mHz from the frequency of the injection.
The injected signal ip6 has a spin-down of $-6.73\times10^{-9}$~Hz/s, which is outside our search range.
However, the SNR of the injection was around 17 for each of the 6-day segments that we analyzed.
This resulted in a sufficiently strong correlation to give a significant signal;
however, with the spin-down and the sky position of the outlier very much displaced from the true values (see Table \ref{tab:tdfstat_ip6}).
\begin{table*}[tbp]
\begin{center}
\begin{tabular}{l c@{\hspace{3ex}} c c@{\hspace{3ex}} c@{\hspace{3ex}} c}
\hline
Injection & FAP & $\Delta f$~[Hz] & $\Delta\dot{f}$ [nHz/s] & $\Delta\delta$ [deg] & $\Delta\alpha$ [deg] \\
\hline \hline
ip6 & $4.02\times10^{-8}$ & $1.54 \times 10^{-2}$ & $2.26$   & 36.59  & 314.54 \\
\hline
\end{tabular}
\caption[]{Outlier associated with the hardware injection ip6.}
\label{tab:tdfstat_ip6}
\end{center}
\end{table*}

The {\bf 102} outliers that could not be associated with interferences in the detector or hardware injections
appeared with frequencies on the left edges of the 0.25~Hz sub-bands of the narrowband segments that we analyzed.
To determine whether these are artifacts or they warrant a further detailed follow-up,
we regenerated the narrowband data where the artefacts occurred,
however with the offset frequencies decreased by 0.125~Hz (half of the width of the sub-band).
Consequently the outliers that appeared at the left edges of the sub-bands, should now be present approximately in the middle of the sub-band.
We have then performed a search with our pipeline around the parameters of the outliers.
None of the outliers were found to be significant. The smallest false probability was found to be around 59\%.   

As a result we were left with {\bf 2} outliers  for a more detailed study, with parameters given in Table \ref{tab:tdfstat_out}.
We followed up the outliers in the data segments that are twice as long as the original segments.
For each sub-band where the outliers are present we divided the data into 12-day segments and we performed the search around the position of the outliers.
A two-fold increase of the coherence times would result in the increase of the signal-to-noise ratio of a true GW signal by a factor of $\sqrt{2}$.
We performed a coherent search $\pm 16$ grid points in spin down and $\pm 4$ points in the sky position around the point of the parameter space where the outliers should be present.
We then performed a coincidence search. For the two cases we did not find a significant coincidence.
The probability that the best coincidence was accidental was close to 1.

\begin{table*}[tbp]
\begin{center}
\begin{tabular}{d D..{4.8} D..{6.4} D..{5.4} D..{1.4}}
\hline
\multicolumn{1}{c}{$f$~\text{[Hz]}} &
\multicolumn{1}{c}{$\dot{f}$~\text{[nHz/s]}} &
\multicolumn{1}{c}{$\delta$~\text{[deg]}} &
\multicolumn{1}{c}{$\alpha$~\text{[deg]}} &
\multicolumn{1}{c}{\text{FAP}}\\
\hline \hline
 83.52 & -6.58\times10^{-1} & -18.08 & 179.16 & 0.0094\\
726.07 & -3.30\times10^{-2} &  58.07 & 190.79 & 0.0034\\
\hline
\end{tabular}
\caption[]{Outliers of unknown origin from the Time-Domain $\F$-statistic analysis.}
\label{tab:tdfstat_out}
\end{center}
\end{table*}

\subsubsection{Upper limits}
\label{ssec:fstat_uls}

\begin{figure}
\centering
\includegraphics[width=\columnwidth]{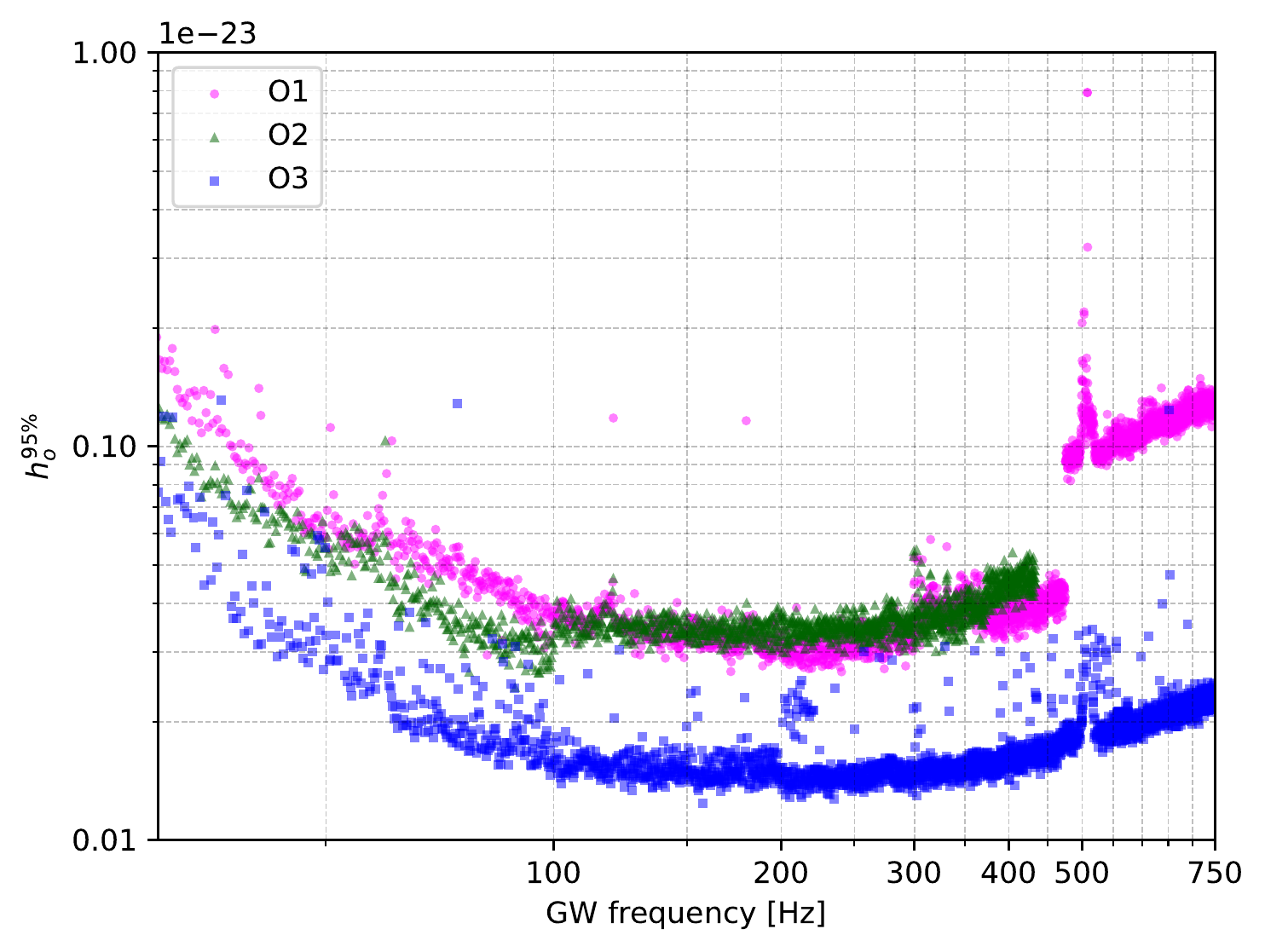}
\caption{Comparison of 95\% confidence upper limits on GW amplitude $h_0$
obtained with the {\tdfstat} pipeline in the analysis of Advanced LIGO data.
The magenta circles, green triangles, and blue squares represent the $h_0^{95\%}$ upper limits in 0.25~Hz sub-bands of the O1, O2, and O3 data, respectively.}
\label{fig:tdfstat_ul_O1O2O3}
\end{figure}

The analysis of the outliers described in Secs.\ \ref{ssec:fstat_HI} and \ref{ssec:fstat_outliers} has not revealed a viable candidate for a GW event.
We therefore proceeded to establish upper limits on the amplitude of GW signals in our search.
We establish upper limits in each sub-band analyzed and for each hemisphere by using the procedure described in Sec.\ \ref{sssec:Fstatsen}
(as a result periodic interferences in the data for 201 sub-bands out of 3141 that we analyzed we were not able to establish upper limits).
The 95\% confidence upper limits $h_0^{95\%}$ for analysis of LIGO O3 data presented in the paper are plotted in Fig.\ \ref{fig:tdfstat_ul_O1O2O3}
in comparison with upper limits obtained with our pipeline in O1 and O2 data.
We see a considerable improvement which is more than the improvement in the sensitivity of LIGO data.
This additional improvement in our pipeline sensitivity is around 1/3 and it is mainly due to changes in the coincidence algorithm.
The biggest improvement for frequencies above 450~Hz is due to the longer coherence time of 6 days used in the search,
compared to the coherence time of 2 days used in our O2 search above 450~Hz.

\subsection{\soap}
\label{ssect:SOAPr}

\soap \, was run on the O3 dataset from 40-2000 Hz where we are sensitive to a broad range of signals from the entire sky. To contain an entire signal within a single sub-band, its spin-down must be within  $\pm \sim 10^{-9}$ Hz/s up to 1000 Hz and $\pm \sim 10^{-8}$ Hz/s above 1000 Hz, therefore when values are outside this range we lose sensitivity.
We start from a set of 1800s long \acp{FFT} of cleaned time-series data from the two LIGO detectors H1 and L1. 
As described in Sec.~\ref{ssect:SOAPm:data}, the \acp{FFT} are normalised to the running median of width 100 bins before being split into 0.1 (0.2, 0.3, 0.4) Hz wide sub-bands overlapping by half of their width.
For each of the sub-bands, time segments and frequency bins are summed together, where along the time axis, 48 \acp{FFT} (1 day) are summed along the frequency axis, and every 1 (2, 3, 4) frequency bins are summed respective to the analysis band. 
\soap \, is then run on each of these sub-bands, returning the Viterbi statistic, Viterbi map and Viterbi tracks, which can be input to the \ac{CNN} to return a second statistic.
The number of sub-bands searched totals to 19 868 across all four analysis bands, where for each band (40-500, 500-1000, 1000-1500, 1500-2000 ) Hz the respective total is (9200, 5040, 3263, 2362).
Sub-bands which contained known instrumental lines identified by the calibration group are then removed from the analysis leaving a total number of sub-bands as 17 929, with each separate band containing (8297, 4494, 2952, 2186) sub-bands.
Candidates are then selected by taking the sub-bands which contribute to the top 1\% of both the remaining Viterbi and \ac{CNN} statistics. 
These candidates can then be investigated further to identify whether a real GW signal is present.
Sub-bands which contain an instrumental line identified by the calibration group but also cross the 1\% threshold are also investigated to check whether it is the instrumental line which causes the high statistic value.
There were 293 sub-bands which were in this category, and in 291 sub-bands the Viterbi tracks closely follow the instrumental line, and the remaining two contained both an instrumental line and a hardware injection (ip5). 
These were then reintroduced into the analysis as the Viterbi tracks did not follow that of the instrumental line.
From the total of the 17 929 sub-bands, 248 were selected for a followup investigation where 107 of these sub-bands cross the thresholds of both the Viterbi and \ac{CNN} statistics.

\subsubsection{Outliers}

The 248 candidates are then investigated further by analysing the outputs of the Viterbi search, i.e. the Viterbi maps, Viterbi tracks and Viterbi statistics, alongside the \ac{CNN} statistic and the spectrograms from each detector.
Plots of each of these allow the identification of features which are not astrophysical but originate from the instrument or environment. 
The spectrograms from both detectors summed over time and frequency, as described in Sec.~\ref{ssect:SOAPm:data}, along with the optimal Viterbi track, allow us to identify what features within the data contribute towards the final statistic. 
For example, many of the spectrograms contain spectral features which are far above the noise level and appear in only a single detector, but still crosses the detection threshold for one of the statistics. 
These sub-bands can be visually inspected, and if found to contain a non-astrophysical artefact which contributes to the statistic is removed from the analysis.
Of the sub-bands that were investigated further, 242 were removed due to the presence of an instrumental artefact.
These range from broad spectral lines which last the entire observing run to short duration ($\mathcal{O}(\rm{days})$) high power events which contribute large amounts of power to the statistic.
The remaining sub-bands contain fake \ac{CW} signals which were injected into the hardware of the detector.

\subsubsection{Hardware injections}

In O3 there are a total of 18 hardware injections, where 9 fall within our search parameter space and two of these (ip1 and ip5) appear in sub-bands which cross our detection threshold without being excluded. 
These signals appear in multiple sub-bands due to the 50\% overlap, therefore the sub-band containing a larger fraction of the signal is used for followup. 
Two additional injections outside of our 'sensitive' range for $\dot{f}$ also crossed our detection thresholds (ip4 and ip6) as \soap \, identified the part of the signal which crossed the search band.
Of the 7 injections we did not detect, two are in binary systems which we are less likely to detect as this search was optimised for isolated NSs.
The remaining missed injections have SNRs which are below our expected sensitivity for isolated NSs, therefore would not be expected to cross our threshold.
The two remaining hardware injections crossed the detection threshold for both the Viterbi statistic or the \ac{CNN} statistic.
These candidates were then followed up using the parameter estimation method described in Sec.~\ref{ssect:SOAPm:parest}, where we correctly recover the injected parameters of the injections.

\subsubsection{Sensitivity}

\begin{figure} 
    \includegraphics[width=\columnwidth]{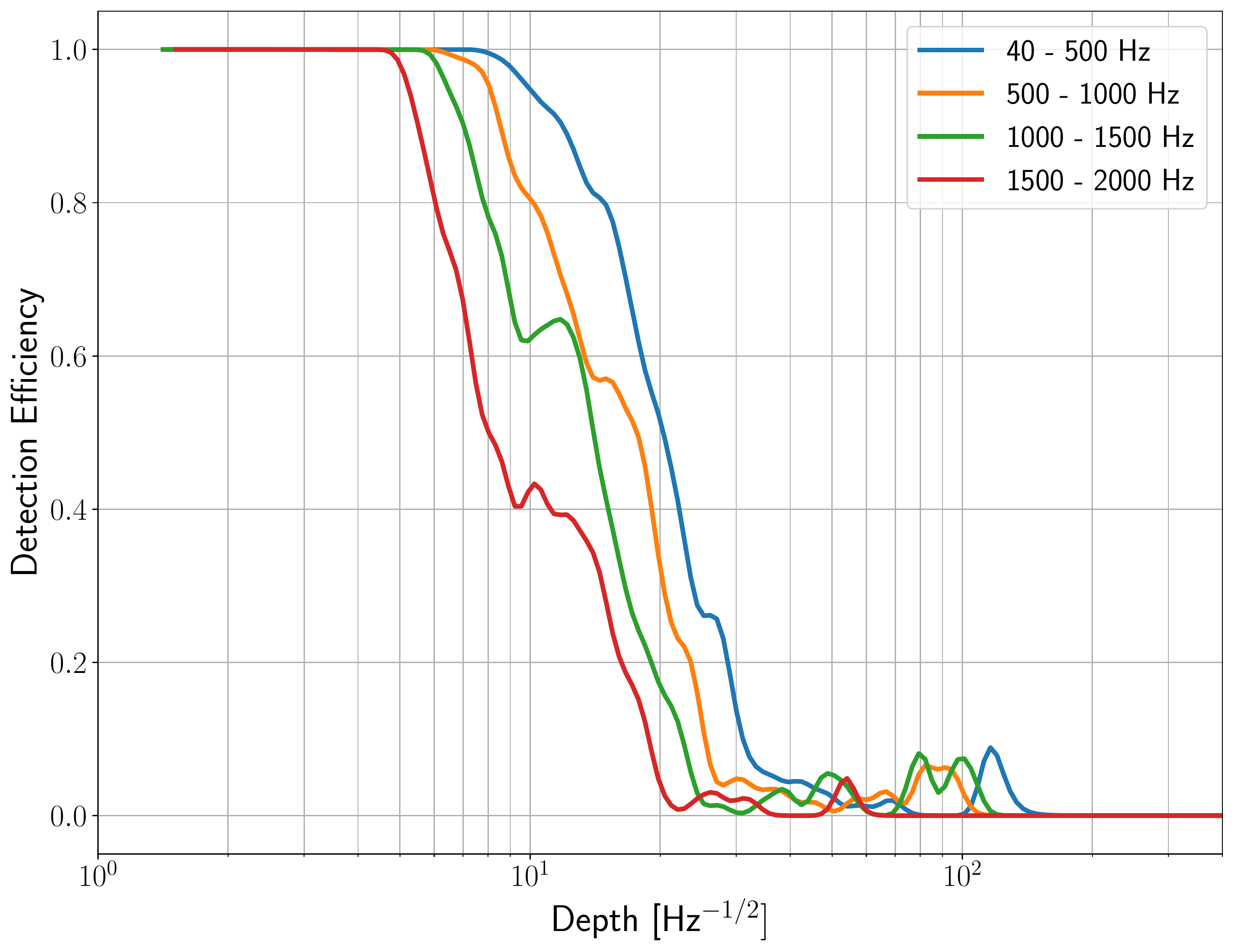}
    \caption{Detection efficiencies of the {\soap} + \ac{CNN} search on isolated NS signals injected into real O3 data. These are shown as a function of the sensitivity depth for the four different frequency ranges described in Sec.~\ref{ssect:SOAPm:data}. The efficiencies are calculated with a false alarm rate of 1\%.} 
    \label{fig:soap:sensitivity} 
\end{figure}

The sensitivity of \soap \, can be tested by running the search on a set of \ac{CW} signals injected into real O3 data. 
A total of $3.3 \times 10^4$ signals are injected across each of the four frequency bands described in Sec.~\ref{ssect:SOAPm:data}, where the signals has Doppler parameters which are drawn uniformly on the sky, uniformly within the respective frequency range and uniformly in the range [$-10^{-9}$, 0] Hz s$^{-1}$ for the frequency derivative. 
The other amplitude parameters varied in the same ranges as described in Sec.~\ref{ssect:SOAPm:CNN}.
A false alarm value of 1\% can be set for each of the odd and even data-sets within the four analysis bands by taking the corresponding statistic value at which 1\% of the noise only bands exceed.
Both the Viterbi and \ac{CNN} statistics are calculated separately for each of the odd and even bands.
Each of the bands containing injected signals can then be classified as detected or not depending on if a statistic crossed its respective false alarm value.
These classified statistics can then be combined together to produce an efficiency curve shown in Fig.~\ref{fig:soap:sensitivity}, which show the fraction of detected signals at a given sensitivity depth, defined in Eq.~\ref{eq:depth}.
At a false alarm value of 1\% and a detection efficiency of 95\% we are sensitive to signals with a depth of 9.9, 8.0, 6.5 and 5.3 Hz$^{-1/2}$ for the frequency bands 40-500, 500-1000, 1000-1500 and 1500-2000 Hz respectively.
To further investigate our sensitivity, we split each of the four analysis bands into smaller bands ranging from 20 Hz wide at lower frequency to 100 Hz wide at higher frequencies.
For each of these bands a detection efficiency curve is generated in the same way as for the sensitivity depth above, however, they are now generated for values of $h_0$. 
The false alarm values for each band are set based on which of the four larger analysis bands that it falls within.
Our false alarm values are then contaminated by the strongest artefacts within each ~500 Hz wide analysis band, meaning that this is a conservative estimate of our sensitivity..
The error on these curves is found using the binomial error on each of the points as defined in Eq.~\ref{binomialerror}, giving two bounds on our efficiency curves.
Values of $h_0$ for each frequency band can then be selected where the detection efficiency reaches 95\%, defining our sensitivity shown in Fig.~\ref{fig:summary-ul}.

\section{Conclusions}
\label{sect:conclusions}

\begin{figure}
    \includegraphics[width=\textwidth]{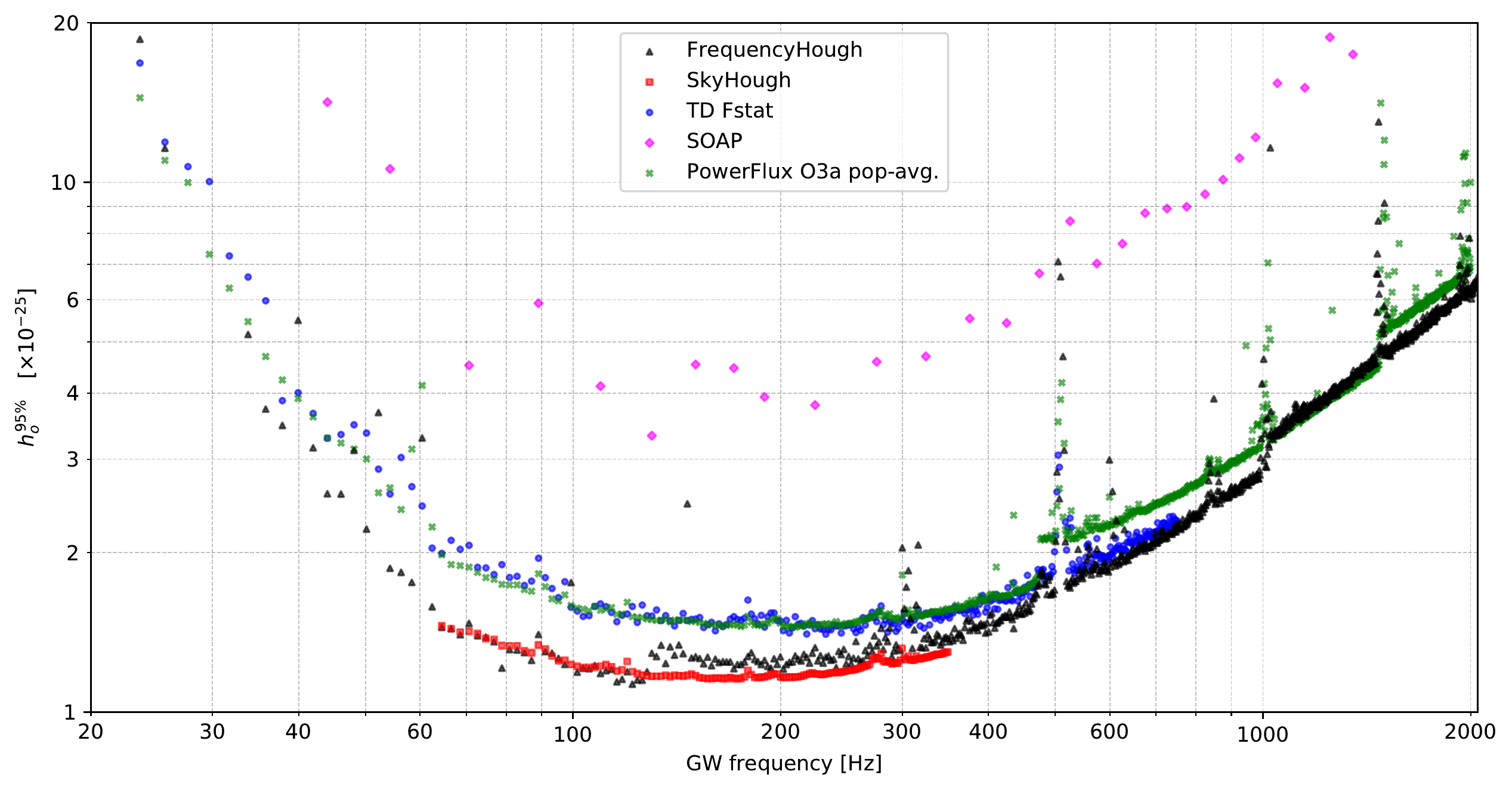}
    \caption{Comparison of 95\% confidence upper limits on GW amplitude $h_0$ obtained by the {\fh} pipeline (black triangles), the {\sh} pipeline (red squares), the {\tdfstat} pipeline (blue circles), and the {\soap} pipeline (magenta diamonds). Population-averaged upper limits obtained in \cite{2021arXiv210700600T} using the O3a data are marked with dark-green crosses. To enhance visibility, we do not show the error estimates of $h_0$ in this plot; additionally, the data is divided in 2 Hz bins, and the median of $h_0$ values within each bin is presented.}
    \label{fig:summary-ul}
\end{figure}

In  Fig.\ \ref{fig:summary-ul} we summarize 95\% confidence-level upper limits on strain amplitude $h_0$ for the four pipelines used in this search. 
The upper limits obtained improve on those obtained using the PowerFlux method
in early O3 LIGO data \cite{PhysRevD.104.082004}.
Our results constitute the most sensitive all-sky search to date
for continuous GWs in the range 20-2000~Hz
while probing spin-down magnitudes as high as $1\times10^{-8}$~Hz/s.
Only the O2 Falcon search \cite{2021PhRvD.103f3019D,2020PhRvL.125q1101D,2021arXiv210409007D}
provides a better sensitivity in the frequency range 20-2000~Hz;
however it does so with a dramatically reduced frequency derivative range.
In the frequency range of [20, 500]~Hz Falcon searches a $\dot{f}$
range from $-3\times10^{-13}$~Hz/s to $3\times10^{-13}$~Hz/s
and $\dot{f}$ range upto [$-7.5\times 10^{-12}$, $3\times 10^{-12}$]~Hz/s for frequencies above 500~Hz.
Thus the Falcon search parameter space is smaller than ours by factor of $\sim 1.8\times 10^4$ below 500~Hz
and factor of $10^3$ above 500~Hz.

\begin{figure}[h]
\includegraphics[width=\textwidth]{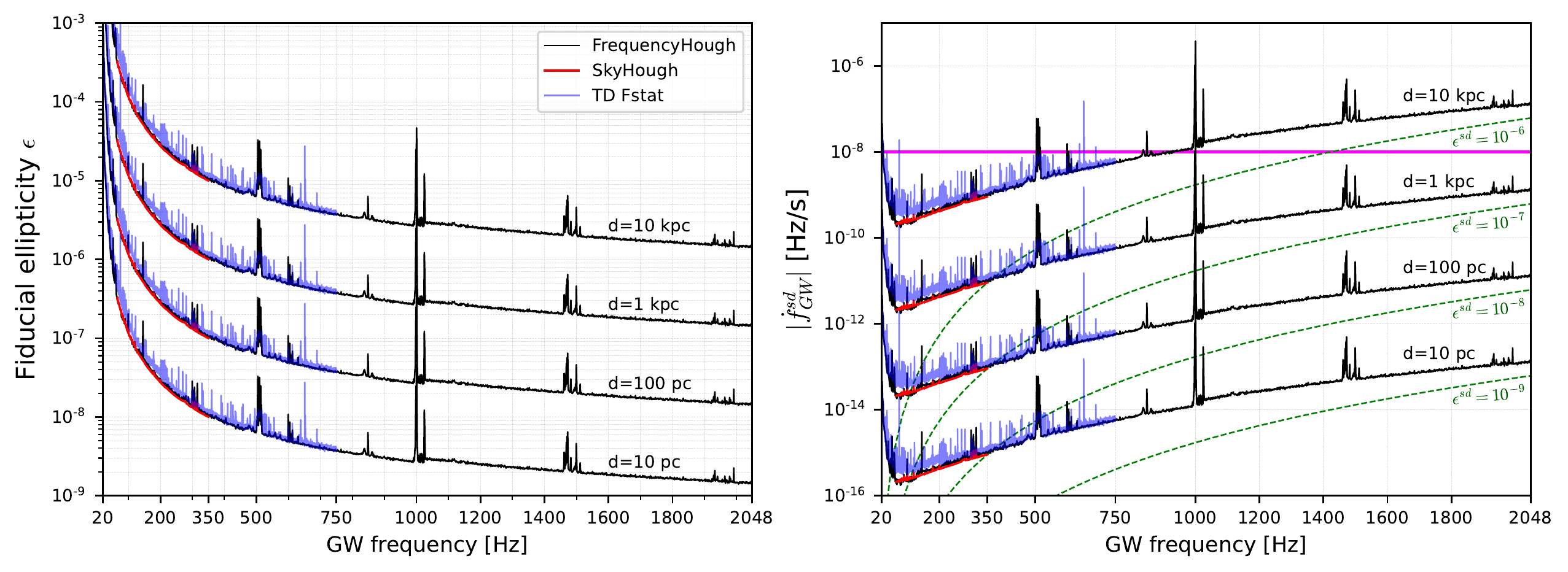}
\caption{Left panel: detectable ellipticity, given by Eq.\ \eqref{eq:fidell}, as a function of the GW frequency for neutron stars with the `canonical' moment of inertia $I_{zz}=10^{38}$~kg~m$^2$ at a distance of 10 kpc, 1 kpc, 100 pc, and 10 pc (from top to bottom). Results for the {\fh} pipeline are marked in black, {\sh} in red and for {\tdfstat} in blue. The right panel shows the relation between the absolute value of the first GW frequency derivative $\dot{f} = 2\dot{f}_{rot}$ and the GW frequency $f = 2f_{rot}$ (with $f_{rot}$ the rotational frequency) of detectable sources as a function of the distance, assuming their spin-down is due solely to the emission of GWs. Constant spin-down ellipticities $\epsilon^{sd}$, corresponding to this condition, are denoted by dashed green curves. The magenta horizontal line marks the maximum spin down searched.}
\label{fig:ell_vs_f}
\end{figure}

We can use the amplitude $h_0$ given by Eq.\ \eqref{eqn:hexpected} to calculate star's ellipticity $\epsilon$,
\begin{align}
\label{eq:fidell}
\epsilon &=
\frac{c^4}{4\pi^2G}\frac{h_0 d}{I_{zz} f^2} \approx \scinum{9.46}{-6}\left(\frac{h_0}{10^{-25}}\right)
\nonumber\\[2ex]
&\qquad\times\left(\frac{10^{38}\,\text{kg}\,\text{m}^2}{I_{zz}}\right)
\left(\frac{100\,\text{Hz}}{f}\right)^2\left(\frac{d}{1~\text{kpc}}\right) .
\end{align}
Using the above equation the upper limits on the GW strain amplitude $h_0$ can be converted to upper limits on the ellipticity $\epsilon$. The results are plotted in Fig.~\ref{fig:ell_vs_f} (left panel) for four representative values of the distance d and they provide astrophysically interesting results. 
The NSs with ellipticities above a given trace and distance value corresponding to the trace in the left panel of Fig.~\ref{fig:ell_vs_f} would be detectable by our searches.
For instance, at frequency 200 Hz we would be able to detect a CW signal from a NS within a distance of 100 pc if its ellipticity were at least
$3\times10^{-7}$. Similarly, in the middle frequency range, around
550 Hz, we would be able to detect the CW signal up
to a distance of 1 kpc, with $\epsilon > 5\times 10^{-7}$. Finally at higher
frequencies, around 1550 Hz, the same signal would be
detectable up to a distance of 10 kpc if $\epsilon > 2\times 10^{-6}$. 
These levels of ellipticity are below the maximum value of the ellipticity that may be supported by the crust of a NS described by a standard equation of state reported in \cite{10.1046/j.1365-8711.2000.03938.x,10.1111/j.1365-2966.2006.10998.x,2013PhRvD..88d4004J}. However they are above the most recent estimates in general relativity by \cite{10.1093/mnras/stab2048,10.1093/mnras/staa3635}. The latter do not, however, exclude
larger values of ellipticity when additional physical processes, such as plastic flow in the crust, are taken into account.
Our upper limits are starting to probe the range predicted for pulsars by the models of \cite{Suvorov2016}, which predict ellipticities up to $\epsilon\approx 10^{-7}-10^{-6}$ for younger stars in which the deformation is not supported by crustal rigidity, but by a non-axisymmetric magnetic field at the end of its Hall driven evolution in the crust. Note however that for known pulsars at a distance of a few kpc, such as the Crab, the signal would be at frequencies $f\lesssim 100$ Hz, so still beyond the reach of our searches. 

Another way of representing limits on ellipticity is shown in the right panel of Fig.\ \ref{fig:ell_vs_f}.
Assuming that the emission of gravitational radiation is the sole energy loss mechanism for a rotating NS,
we obtain the so-called spin-down limit $h_0^{\text{sd}}$ on the amplitude $h_0$, see Eqs.\ (7)--(9) of \cite{2019ApJ...879...10A}:
\begin{align}
\label{eq:h0sd}
h_0^{\text{sd}} &= \frac{1}{d}\left(\frac{5}{2} \frac{G I_{zz}}{c^3}\frac{|\dot{f}|}{f}\right)^{1/2}
\approx \scinum{2.55}{-25} \left(\frac{1~\text{kpc}}{d}\right)
\nonumber\\
&\quad\times\left(\frac{I_{zz}}{10^{38}\,\text{kg}\,\text{m}^2}\right)^{1/2}
\left(\frac{100\,\text{Hz}}{f}\right)^{1/2}
\left(\frac{|\dot{f}|}{10^{-11}~\text{Hz}\,\text{s}^{-1}} \right)^{1/2}.
\end{align}
Inverting the above equation and replacing the spin-down limit amplitude $h_0^{\text{sd}}$ with our upper limit amplitudes $h_0^{95\%}$ we have the following relation between the frequency derivative and frequency:
\begin{align}
\label{eq:fdot_ul}
|\dot{f}| &= \frac{2 c^3}{5 G} \frac{(h_0^{95\%} d)^2 f}{I_{zz}} \approx \scinum{1.54}{-10}\left(\frac{h_0^{95\%}}{10^{-24}}\right)^2
\nonumber\\
&\quad\times\left(\frac{10^{38}\,\text{kg}\,\text{m}^2}{I_{zz}}\right)
\left( \frac{f}{100\,\text{Hz}}\right) \left(\frac{d}{1\,\text{kpc}}\right)^2.
\end{align}

In the right panel of Fig.\ \ref{fig:ell_vs_f} we have plotted $|\dot{f}|$ as a function of frequency $f$
for several representative values of the distance $d$ and for a canonical value of the moment of inertia.
The NSs with $|\dot{f}|$ above a given trace and distance value corresponding to the trace in the right panel of Fig.~\ref{fig:ell_vs_f} would be detectable by our searches.

By equating Eq.\ \eqref{eqn:hexpected} for the amplitude $h_0$ and Eq.\ \eqref{eq:h0sd} for the spin-down limit,
we obtain the following equation for $\dot{f}$:
\begin{align}
\label{eq:fdot_sd}
|\dot{f}| &= \frac{32 \pi^4 G}{5 c^5} \epsilon^2 I_{zz} f^5 \approx \scinum{1.72}{-14}
\left(\frac{\epsilon}{10^{-6}}\right)^2
\nonumber\\
&\qquad \times \left(\frac{I_{zz}}{10^{38}\,\text{kg}\,\text{m}^2}\right)\left( \frac{f}{100\,\text{Hz}}\right)^5.
\end{align}
The dashed lines in the right panel of Fig.\ \ref{fig:ell_vs_f} are constant ellipticity curves from Eq.~\eqref{eq:fdot_sd} above. These lines are independent of the distance $d$. 

\begin{figure} 
\includegraphics[width=\columnwidth]{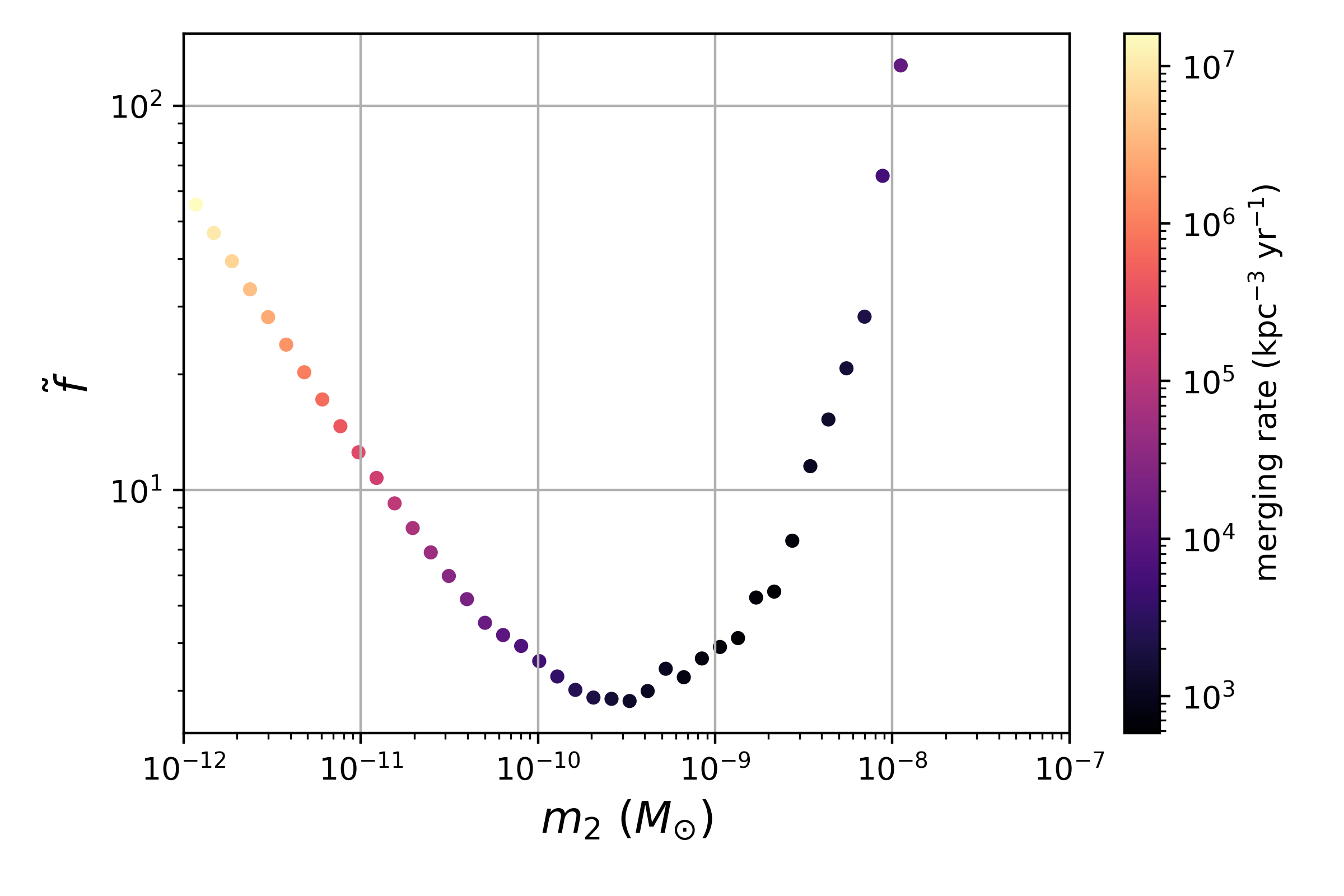}
\caption{Constraints on $\tilde{f}$, a quantity that, if less than one, indicates the sensitivity to a given $f_{\rm pbh}$, and inspiraling rate (color) as a function of the secondary mass, with a primary mass $m_1=2.5M_\odot$, assuming a monochromatic mass function for $m_1$, no rate suppression, and $f_{\rm pbh}=1$. These constraints are valid at distances of $\mathcal{O}$(pc).} 
\label{fig:pbhcw} 
\end{figure}

In addition to constraints on ellipticities of isolated Ns, we can make statements about the rate and abundance of inspiraling planetary-mass and asteroid-mass PBHs \cite{Miller:2020kmv}. The upper limits presented in Fig.~\ref{fig:summary-ul} are \emph{generic}: they can be applied to any quasi-monochromatic, persistent GW that follows a linear frequency evolution over time and whose frequency derivative lies within the search range. Based on these all-sky searches, GW signals from inspiralling PBH binaries with chirp masses less than $\mathcal{O}(10^{-5})M_\odot$ and GW frequencies less than $\sim 250$ Hz would be identical to those arising from non-axisymmetric rotating NSs. Following the procedure presented in \cite{Miller:2021knj}, and using the \fh\ upper limits in Fig.~\ref{fig:fhulresults}, which cover the widest range of spin-down/spin-up, we obtain constraints on highly asymmetric mass ratio binary systems, assuming that one object in the binary has a mass $m_1=2.5M_\odot$, motivated by the QCD phase transition \cite{Byrnes:2018clq,Carr:2019kxo,Clesse:2020ghq}. In Fig.~\ref{fig:pbhcw}, we plot constraints on the merging rates and an effective parameter, $\tilde{f}$, that, if less than one, indicates the sensitivity to the fraction of dark matter that PBHs could compose, $f_{\rm pbh}$, as a function of the companion mass $m_2$:

\begin{equation}
    \tilde f^{53/37} \equiv f_{\rm sup} f(m_1) f(m_2) f_{\rm pbh}^{53/37},
\end{equation}
where $f_{\rm sup}$ is a rate suppression factor, defined to be one, and $f(m_1)$ and $f(m_2)$ are the mass distribution functions for $m_1$ and $m_2$, respectively. We assume $f(m_1)=1$, that is, a monochromatic mass function peaked at 2.5$M_\odot$, and that $f_{\rm pbh}=1$. Our results indicate that $\tilde{f}$ lies slightly above one for a wide range of asteroid-mass BHs within the distance range of $\mathcal{O}$(pc). However, as the GW detectors become more sensitive, and especially when future ground-based instruments come online, we will start to probe a physical regime of the PBH masses. Thus, these results not only imply a bright future for analyses for PBHs, but also motivate the expansion of CW techniques specifically adapted to search for planetary-mass inspiraling PBHs \cite{Miller:2020kmv}. While the rates only depend on the distance reach of the search, i.e. they are model-independent, the constraints on $\tilde{f}$ depend on particular models of PBH clustering or binary formation. Other models are certainly just as valid as those that we constrain here \cite{raidal2019formation,Hutsi:2020sol}; therefore, Fig.~\ref{fig:pbhcw} should be seen as an example of the kinds of statements that could be made based on upper limits from CW searches, and not an absolute statement about the abundance and rates of PBHs.

\newpage 
\onecolumngrid 

\twocolumngrid 

\begin{acknowledgments}

\newif\ifcoreonly\coreonlyfalse
\newif\ifkagra\kagratrue
\newif\ifheader\headerfalse
\ifheader
\begin{center}{\bf\Large
\ifkagra
Conference proceedings acknowledgements for \\ the LIGO Scientific Collaboration, the Virgo Collaboration and the KAGRA Collaboration
\else
Conference proceedings acknowledgements for \\ the LIGO Scientific Collaboration and the Virgo Collaboration
\fi
}\end{center}
\fi
This material is based upon work supported by NSF’s LIGO Laboratory which is a major facility
fully funded by the National Science Foundation.
The authors also gratefully acknowledge the support of
the Science and Technology Facilities Council (STFC) of the
United Kingdom, the Max-Planck-Society (MPS), and the State of
Niedersachsen/Germany for support of the construction of Advanced LIGO 
and construction and operation of the GEO\,600 detector. 
Additional support for Advanced LIGO was provided by the Australian Research Council.
The authors gratefully acknowledge the Italian Istituto Nazionale di Fisica Nucleare (INFN),  
the French Centre National de la Recherche Scientifique (CNRS) and
the Netherlands Organization for Scientific Research (NWO), 
for the construction and operation of the Virgo detector
and the creation and support  of the EGO consortium. 
The authors also gratefully acknowledge research support from these agencies as well as by 
the Council of Scientific and Industrial Research of India, 
the Department of Science and Technology, India,
the Science \& Engineering Research Board (SERB), India,
the Ministry of Human Resource Development, India,
the Spanish Agencia Estatal de Investigaci\'on (AEI),
the Spanish Ministerio de Ciencia e Innovaci\'on and Ministerio de Universidades,
the Conselleria de Fons Europeus, Universitat i Cultura and the Direcci\'o General de Pol\'{\i}tica Universitaria i Recerca del Govern de les Illes Balears,
the Conselleria d'Innovaci\'o, Universitats, Ci\`encia i Societat Digital de la Generalitat Valenciana and
the CERCA Programme Generalitat de Catalunya, Spain,
the National Science Centre of Poland and the European Union – European Regional Development Fund; Foundation for Polish Science (FNP),
the Swiss National Science Foundation (SNSF),
the Russian Foundation for Basic Research, 
the Russian Science Foundation,
the European Commission,
the European Social Funds (ESF),
the European Regional Development Funds (ERDF),
the Royal Society, 
the Scottish Funding Council, 
the Scottish Universities Physics Alliance, 
the Hungarian Scientific Research Fund (OTKA),
the French Lyon Institute of Origins (LIO),
the Belgian Fonds de la Recherche Scientifique (FRS-FNRS), 
Actions de Recherche Concertées (ARC) and
Fonds Wetenschappelijk Onderzoek – Vlaanderen (FWO), Belgium,
the Paris \^{I}le-de-France Region, 
the National Research, Development and Innovation Office Hungary (NKFIH), 
the National Research Foundation of Korea,
the Natural Science and Engineering Research Council Canada,
Canadian Foundation for Innovation (CFI),
the Brazilian Ministry of Science, Technology, and Innovations,
the International Center for Theoretical Physics South American Institute for Fundamental Research (ICTP-SAIFR), 
the Research Grants Council of Hong Kong,
the National Natural Science Foundation of China (NSFC),
the Leverhulme Trust, 
the Research Corporation, 
the Ministry of Science and Technology (MOST), Taiwan,
the United States Department of Energy,
and
the Kavli Foundation.
The authors gratefully acknowledge the support of the NSF, STFC, INFN, CNRS and PL-Grid for provision of computational resources.

%\ifcoreonly\else
%{\bf For papers using O3b (and future) data, the following paragraph should be added for KAGRA.}\\
%\fi
\ifkagra
% updated June 2021
This work was supported by MEXT, JSPS Leading-edge Research Infrastructure Program, JSPS Grant-in-Aid for Specially Promoted Research 26000005, JSPS Grant-in-Aid for Scientific Research on Innovative Areas 2905: JP17H06358, JP17H06361 and JP17H06364, JSPS Core-to-Core Program A. Advanced Research Networks, JSPS Grant-in-Aid for Scientific Research (S) 17H06133 and 20H05639 , JSPS Grant-in-Aid for Transformative Research Areas (A) 20A203: JP20H05854, the joint research program of the Institute for Cosmic Ray Research, University of Tokyo, National Research Foundation (NRF), Computing Infrastructure Project of KISTI-GSDC, Korea Astronomy and Space Science Institute (KASI), and Ministry of Science and ICT (MSIT) in Korea, Academia Sinica (AS), AS Grid Center (ASGC) and the Ministry of Science and Technology (MoST) in Taiwan under grants including AS-CDA-105-M06, Advanced Technology Center (ATC) of NAOJ, and Mechanical Engineering Center of KEK.
\fi

%\ifcoreonly\else
%{\bf For certain collaboration papers, it may be appropriate to acknowledge specific analysis software.
%One template to consider is that used for the GW190425 discovery paper, for which the latex can be
%found here:}\\
%{\small https://git.ligo.org/publications/gw190425/gw190425-discovery/-/blob/master/gw190425-discovery.tex\#L259.}
%\fi

%\ifcoreonly\else
%{\bf For collaboration papers released after March 2020, it may be appropriate to add the following special acknowledgement, BUT the journal may reject it (PRL rejected it). If one chooses, one can include such acknowledgements in arXiv/DCC versions only in that case:}\\
%\fi
\ifcoreonly\else
{\it We would like to thank all of the essential workers who put their health at risk during the COVID-19 pandemic, without whom we would not have been able to complete this work.}
\fi

\end{acknowledgments}

\bibliographystyle{unsrt}
\bibliography{references} 

\iftoggle{endauthorlist}{
  \let\author\myauthor
  \let\affiliation\myaffiliation
  \let\maketitle\mymaketitle
  \title{The LIGO Scientific Collaboration, Virgo Collaboration, and KAGRA Collaboration}
  \pacs{}

%% LVK authorlist in PRD format
% \documentclass[aps,prd,superscriptaddress]{revtex4-2}
% \usepackage{orcidlink}
% \begin{document}

% \title{LSC, Virgo and KAGRA August 2021 author list---LIGO-M2100182\\
% 2021-10-20. PRD style}

\author{R.~Abbott}
\affiliation{LIGO Laboratory, California Institute of Technology, Pasadena, CA 91125, USA}
\author{H.~Abe}
\affiliation{Graduate School of Science, Tokyo Institute of Technology, Meguro-ku, Tokyo 152-8551, Japan  }
\author{F.~Acernese}
\affiliation{Dipartimento di Farmacia, Universit\`a di Salerno, I-84084 Fisciano, Salerno, Italy  }
\affiliation{INFN, Sezione di Napoli, Complesso Universitario di Monte S. Angelo, I-80126 Napoli, Italy  }
\author{K.~Ackley\,\orcidlink{0000-0002-8648-0767}}
\affiliation{OzGrav, School of Physics \& Astronomy, Monash University, Clayton 3800, Victoria, Australia}
\author{N.~Adhikari\,\orcidlink{0000-0002-4559-8427}}
\affiliation{University of Wisconsin-Milwaukee, Milwaukee, WI 53201, USA}
\author{R.~X.~Adhikari\,\orcidlink{0000-0002-5731-5076}}
\affiliation{LIGO Laboratory, California Institute of Technology, Pasadena, CA 91125, USA}
\author{V.~K.~Adkins}
\affiliation{Louisiana State University, Baton Rouge, LA 70803, USA}
\author{V.~B.~Adya}
\affiliation{OzGrav, Australian National University, Canberra, Australian Capital Territory 0200, Australia}
\author{C.~Affeldt}
\affiliation{Max Planck Institute for Gravitational Physics (Albert Einstein Institute), D-30167 Hannover, Germany}
\affiliation{Leibniz Universit\"at Hannover, D-30167 Hannover, Germany}
\author{D.~Agarwal}
\affiliation{Inter-University Centre for Astronomy and Astrophysics, Pune 411007, India}
% ...

\collaboration{The LIGO Scientific Collaboration, the Virgo Collaboration, and the KAGRA Collaboration}

  \newpage
  \maketitle
}

\end{document}